\documentclass[12pt]{emulateapj}
\usepackage{ulem}
\slugcomment{To be submitted to the Astrophysical Journal}

\begin{document}

\title{Interacting Binaries with Eccentric Orbits.  Secular 
Orbital Evolution Due To Conservative Mass Transfer} 

\author{J. F. Sepinsky, B. Willems, V. Kalogera, F. A. Rasio}
\affil{Department of Physics and Astronomy, Northwestern University,
2145 Sheridan Road, Evanston, IL 60208}
\slugcomment{j-sepinsky, b-willems, vicky@northwestern.edu, and
rasio@northwestern.edu}
\shortauthors{Sepinsky, Willems, Kalogera, \& Rasio}
\shorttitle{Interacting Binaries with Eccentric Orbits}

\begin{abstract}

We investigate the secular evolution of the orbital semi-major axis and
eccentricity due to mass transfer in eccentric binaries, assuming
conservation of total system mass and orbital angular momentum. Assuming
a delta function mass transfer rate centered at periastron, we find
rates of secular change of the orbital semi-major axis and eccentricity
which are linearly proportional to the magnitude of the mass transfer
rate at periastron. The rates can be positive as well as negative, so
that the semi-major axis and eccentricity can increase as well as
decrease in time. Adopting a delta-function mass-transfer rate of
$10^{-9} M_\sun\,{\rm yr}^{-1}$ at periastron yields orbital evolution
timescales ranging from a few Myr to a Hubble time or more, depending on
the binary mass ratio and orbital eccentricity. Comparison with orbital
evolution timescales due to dissipative tides furthermore shows that
tides cannot, in all cases, circularize the orbit rapidly enough to
justify the often adopted assumption of instantaneous circularization at
the onset of mass transfer.  The formalism presented can be incorporated
in binary evolution and population synthesis codes to create a
self-consistent treatment of mass transfer in eccentric binaries.

\end{abstract}

\keywords{Celestial mechanics, Stars: Binaries: Close, Stars: Mass Loss}

\section{Introduction}
\label{sec-intro}
Mass transfer between components of close binaries is a common
evolutionary phase for many astrophysically interesting binary
systems.  Indeed, mass ejection and/or accretion is responsible for
many of the most recognizable phenomena associated with close
binaries, such as persistent or transient X-ray emission, neutron star
spin-up, and orbital contraction or expansion.  Theoretical
considerations of these and other associated phenomena in the
literature probe these systems quite effectively, yet they often do
not consider the effects of any eccentricity associated with the
binary orbit.  This can be of particular importance for binaries
containing a neutron star or a black hole, where mass loss and natal
kicks occurring during compact object formation may induce a
significant eccentricity to the binary \citep[e.g.][]{H83, BP95,
  K96}. After the formation of the compact object, tides tend to
circularize the orbit on a timescale which strongly depends on the
ratio of the radius of the compact object's companion to the orbital
semi-major axis.  Because of this, orbits are usually assumed to
circularize instantaneously when a binary approaches or begins a mass
transfer phase. 

Despite our generally well developed understanding of tidal
interactions in close binaries, quantitative uncertainties in tidal
dissipation mechanisms propagate into the determination of
circularization timescales. For example, \citet{MM05} have shown
that current theories of tidal circularization cannot explain observed
degrees of circularization of solar-type binaries in open clusters. 
Circularization of high-mass binaries, on the other hand, is currently
thought to be driven predominantly by resonances between dynamic tides
and free oscillation modes, but initial conditions play an
important role and an extensive computational survey of relevant parts
of the initial parameter space has yet to be undertaken \citep{WS99,
WS01, WVS03}.

Furthermore, assumptions of instantaneous circularization immediately
before or at the onset of mass transfer are in clear contrast with
observations of eccentric mass transferring systems. In the most recent
catalog of eccentric binaries with known apsidal-motion rates compiled
by \citet{PO99}, 26 out of the 128 listed systems are semi-detached or
contact binaries. Among these mass-transferring systems, $9$ have
measured eccentricities greater than $0.1$. In addition, many
high-mass X-ray binaries are known to have considerable orbital
eccentricities \citep{Rag05}. While mass transfer in these systems is
generally thought to be driven by the stellar wind of a massive O- or
B-star, it has been suggested that some of them may also be subjected
to atmospheric Roche-lobe overflow at each periastron passage of the
massive donor \citep[e.g.][]{P78}. 

\citet{H56}, \citet{K64-2}, and \citet{P64} were the first to study
the effects of mass transfer on the orbital elements of eccentric
binaries. However, their treatment was restricted to perturbations of
the orbital motion caused by the variable component masses.
\citet{MW83,MW84} extended these early pioneering studies to include
the effects of linear momentum transport from one star to the other,
as well as any other possible perturbations caused by the mass
transfer stream in the system. However, these authors derived the
equations governing the motion of the binary components with respect
to a reference frame with origin at the mass center of the binary,
which is not an inertial frame. Their equations therefore do not
account for the accelerations of the binary mass center caused by the
mass transfer (see \S~\ref{sec-appb}). 

More recent work on mass transfer in eccentric binaries has mainly
focused on smoothed particle hydrodynamics calculations of the mass
transfer stream over the course of a few orbits, without any
consideration of the long-term evolution of the binary \citep{L98,
Rea05}.

Hence, there is ample observational and theoretical motivation to
revisit the study of eccentric mass-transferring binaries.  In this
paper, our aim is to derive the equations governing the evolution of
the orbital semi-major axis and eccentricity in eccentric
mass-transferring binaries, assuming conservation of total system mass
and orbital angular momentum. In a subsequent paper, we will
incorporate the effects of mass and orbital angular momentum losses
from the system.

Our analysis is based on the seminal work of \citet{H69} who was the
first to derive the equations of motion of the components of eccentric
mass-transferring binaries while properly accounting for the effects
of the variable component masses on the stars' mutual gravitational
attraction, the transport of linear momentum from one star to the
other, the accelerations of the binary mass center due to the
redistribution of mass in the system, and the perturbations of the
orbital motion caused by the mass-transfer stream.  While the
equations of motion derived by \citet{H69} are valid for orbits of
arbitrary eccentricity, the author restricted the derivation of the
equations governing the evolution of the semi-major axis and
eccentricity to orbits with small initial eccentricities.

The paper is organized as follows. In \S\,2 and \S\,3 we present the
basic assumptions relevant to the investigation and derive the equations
governing the motion of the components of an eccentric mass-transferring
binary under the assumption of conservative mass transfer.  The
associated equations governing the rates of change of the semi-major
axis and the orbital eccentricity are derived in \S\,4, while numerical
results for the timescales of orbital evolution due to mass transfer as
a function of the initial binary mass ratio and orbital eccentricity are
presented in \S\,5. For comparison, timescales of orbital evolution due
to dissipative tidal interactions between the binary components are
presented in \S\,6.  \S\,7 is devoted to a summary of our main results
and a discussion of future work. In the appendices, lastly, we derive
an equation for the position of the inner Lagrangian point in eccentric
binaries with non-synchronously rotating component stars
(Appendix~\ref{sec-appa}), and present an
alternative derivation for the equations governing the secular evolution
of the orbital semi-major axis and eccentricity assuming instantaneous
mass transfer between two point masses (Appendix~\ref{sec-appc}).

\section{Basic Assumptions}

We consider a binary system consisting of two stars in an eccentric
orbit with period $P_{\rm orb}$, semi-major axis $a$, and
eccentricity $e$. We let the component stars rotate with angular
velocities $\vec{\Omega}_1$ and $\vec{\Omega}_2$ parallel to the
orbital angular velocity $\vec{\Omega}_{\rm orb}$, and assume the
rotation rates to be uniform throughout the stars. We also note that
the magnitude of $\vec{\Omega}_{\rm orb}$ varies periodically in time
for eccentric binaries, but its direction remains fixed in
space. Because of this, the stars cannot be synchronized with the
orbital motion at all times.

At some time $t$, one of the stars is assumed to fill its Roche lobe
and begins transferring mass to its companion through the inner
Lagrangian point $L_1$. We assume this point to lie on the line
connecting the mass centers of the stars, even though non-synchronous
rotation may cause it to oscillate in the direction perpendicular to
the orbital plane with an amplitude proportional to the degree of
asynchronism \citep{MW83}. Since the donor's rotation axis is assumed
to be parallel to the orbital angular velocity, we can safely assume 
that the transferred mass remains confined to the orbital plane.

We furthermore assume that all mass lost from the donor is accreted by
its companion, and that any orbital angular momentum transported by
the transferred mass is immediately returned to the orbit. The mass
transfer thus conserves both the total system mass and the orbital
angular momentum.

We also neglect any perturbations to the orbital motion other than
those due to mass transfer. At the lowest order of approximation,
these additional perturbations (e.g., due to tides, magnetic breaking,
or gravitational radiation) are decoupled from those due to mass
transfer, and can thus simply be added to obtain the total rates of
secular change of the orbital elements.

\section{Equations of Motion}
\label{sec-eom}

\subsection{Absolute Motion of the Binary Components}

\label{sec-pert}

\begin{figure*}
\epsscale{.9}
\plotone{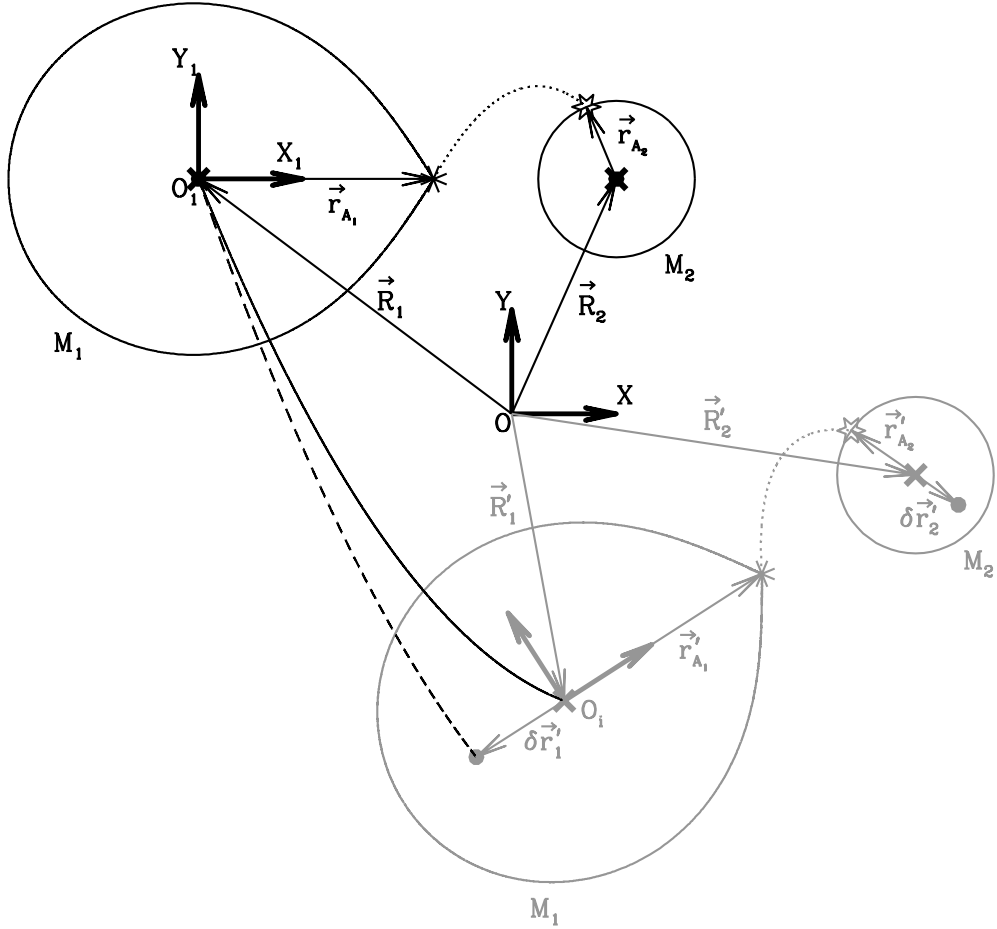}
\caption{
Schematic representation of the reference frames and position vectors
adopted in the derivation of the equations of motion of the 
components of an eccentric mass-transferring binary.  The $Z$- and
$Z_1$-axes of the $OXYZ$ and $O_1X_1Y_1Z_1$ frames are perpendicular
to the plane of the page and are therefore not drawn. The geometry of
the system at time $t$ is shown in black, while the geometry at time
$t + \delta t$ is shown in gray.  The solid line connecting the origin
of the $O_1X_1Y_1Z_1$ frame at time $t$ to the origin of the frame at
time $t+\delta t$ represents the path the donor would have taken had
no mass transfer occurred.  The dashed line, on the other hand,
represents the perturbed motion of the donor's mass center (small
solid circles) due to the mass transfer. A similar perturbation is
imparted to the motion of the accretor. For clarity, this perturbed
motion and the $O_2X_2Y_2Z_2$ reference frame connected to the
accretor (see text) are omitted from the figure. The dotted line 
illustrates a possible path of a mass element transferred from the
donor to the accretor.  The element leaves the donor at the inner
Lagrangian point $L_1$ (asterisk) and accretes onto the companion at
the point $A_2$ (open star).
}
\label{fig-coords}
\end{figure*}

Following \citet{H69}, we derive the equations of motion of the
components of an eccentric mass-transferring binary with respect to a
right handed inertial frame of reference $OXYZ$ which has an arbitrary
position and orientation in space (see Fig.~\ref{fig-coords}). We let
$M_i$ be the mass of star $i$ at some time $t$ at which mass is
transferred from the donor to the accretor, and $M_i + \delta M_i$ the
mass of the same star at some time $t+\delta t$, where $\delta t > 0$ is
a small time interval. With these notations, $\delta M_i < 0$
corresponds to mass loss, and $\delta M_i > 0$ to mass accretion. We
furthermore denote the point on the stellar surface at which mass is
lost or accreted by $A_i$. For the donor star, $A_i$ corresponds to the
inner Lagrangian point $L_1$, while for the accretor, $A_i$ can be any
point on the star's equator. For the remainder of the paper, we let
$i=1$ correspond to the donor and $i=2$ to the accretor.

Because of the mass loss/gain, the center of mass of star $i$ at time
$t + \delta t$ is shifted from where it would have been had no mass
transfer taken place.  To describe this perturbation, we introduce an
additional right-handed coordinate frame $O_i X_i Y_i Z_i$ with a
spatial velocity such that its origin follows the unperturbed orbit of
star $i$, i.e., the origin of $O_iX_iY_iZ_i$ follows the path the
center of mass of star $i$ would have taken had no mass transfer
occurred.  Thus, at time $t$, the center of mass of star $i$ lies at the
origin of $O_i$, while at time $t+\delta t$ it has a non-zero position
vector with respect to $O_i$.  We furthermore let the $Z_i$-axis of the
$O_i X_i Y_i Z_i$ frame point in the direction of the orbital angular
momentum vector, and let the frame rotate synchronously with the 
unperturbed orbital angular velocity of the binary in the absence of
mass transfer. The direction and orientation of the $X_i$-axes are
then chosen such that at time $t$ the $X_i$-axis points along the
direction from the mass center of star $i$ to the mass center of its
companion.

To describe the shift in the mass center of star $i$ due to the mass
loss/gain, we denote the position vector of $O_i$ at times $t$ and $t
+ \delta t$ with respect to the inertial frame by $\vec{R}_{_i}$ and
$\vec{R}_i^\prime$, respectively.  The position vector of the center
of mass of star $i$ at times $t$ and $t + \delta t$ is then given by
$\vec{R}_i$ and $\vec{R}_i^\prime + \delta \vec{r}_i^{\,\prime}$, where 
$\delta \vec{r}_i^{\,\prime}$ is the position vector of the
center of mass of star $i$ at time $t+\delta t$ with respect to
$O_i$.  Moreover, we denote by $\vec{r}_{A_i}$ 
and $\vec{r}_{A_i}^{\,\prime}$
the vectors from $O_i$ to the position where the point $A_i$ on the
stellar surface would be at times $t$ and $t + \delta t$, respectively,
had no mass been lost/accreted. The various position vectors at 
time $t + \delta t$ are related by
\begin{equation}
\left( M_i+\delta M_i \right) \left( \vec{R}_i^\prime 
  + \delta \vec{r}_i^{\,\prime} \right) = M_i\, \vec{R}_i^\prime 
  + \delta M_i\,\left( \vec{R}_i^\prime + \vec{r}_{A_i}^{\,\prime} 
  \right),
\label{eq-cm}
\end{equation}
which, at the lowest order of approximation in $\delta M_i$ and
$\delta \vec{r}_i^{\,\prime}$, yields
\begin{equation}
\delta \vec{r}_i^{\,\prime} = \frac{\delta M_i}{M_i}\, 
  \vec{r}_{A_i}^{\,\prime}.
\label{eq-DelR}
\end{equation}
As expected, the displacement of the center of mass
of star $i$ due to the mass loss/gain is directed along the line
connecting the center of mass of the star and the mass
ejection/accretion point.  

We furthermore denote with $\vec{\rho}_{A_i}^{\, \prime}$ the vector
from the center of mass of star $i$ at time $t+\delta t$ to the
position where the point $A_i$ would be at time $t+\delta t$, had no
mass been transferred between the binary components, and with 
$\delta\vec{\rho}^{\,\prime}_{A_i}$ the perturbation of this vector 
caused by the mass transfer.  It then follows
that $\vec{\rho}_{A_i}^{\, \prime} = \vec{r}_{A_i}^{\,\prime} - \delta
\vec{r}_i^{\,\prime}$ and thus, by definition, $\delta
\vec{\rho}_{A_i}^{\, \prime} = \delta \vec{r}_i^{\,\prime}$.  At the
lowest order of approximation in $\delta M_i$ and $\delta
\vec{r}_i^{\,\prime}$, equation~(\ref{eq-DelR}) therefore also yields
\begin{equation}
\delta \vec{\rho}_{A_i}^{\, \prime} = \frac{\delta
  M_i}{M_i}\, \vec{r}_{A_i}^{\, \prime}.
\label{eq-Delrho}
\end{equation}
The definitions of and the relations between these various position
vectors are illustrated schematically in Fig.~\ref{fig-coords}.

Next, we denote the absolute velocity of the center of mass of star
$i$ with respect to the inertial frame of reference at times $t$ and
$t + \delta t$ by $\vec{V}_i$ and $\vec{V}_i^\prime$, respectively,
and the absolute velocity of the ejected/accreted mass element by 
$\vec{W}_{\delta M_i}$.  The linear momentum $\vec{Q}_1$ of star~$1$ at 
time $t$ is then given by
\begin{equation}
\vec{Q}_1 = M_1\,\vec{V}_1,
\label{eq-Q1-1}
\end{equation}
and the total linear momentum $\vec{Q}_1^\prime$ of star~1 and the
ejected mass element at time $t+ \delta t$ by
\begin{equation}
\vec{Q}_1^\prime = (M_1 + \delta M_1) \vec{V}_1^\prime
  - \delta M_1\,\vec{W}_{\delta M_1}.
\label{eq-Q1-2}
\end{equation}
Similarly, the total linear momentum $\vec{Q}_2$ of star~2 and 
the mass element to be accreted at time $t$ is given by
\begin{equation}
\label{eq-Q2-1}
\vec{Q}_2 = M_2\,\vec{V}_2 + \delta M_i\,\vec{W}_{\delta M_i},
\end{equation}
and the linear momentum $\vec{Q}_2^\prime$ of star~2 at time $t+\delta 
t$ by
\begin{equation}
\label{eq-Q2-2}
\vec{Q}_2^\prime = (M_2 + \delta M_2) \vec{V}_2^\prime.
\end{equation}

At time $t+\delta t$ the velocity of the center of mass of star $i$
can be written as
\begin{equation}
\vec{V}_i^\prime = \vec{V}_{O_i}^\prime
   + \left( \vec{\Omega}_{\rm orb}^\prime 
   + \delta \vec{\Omega}_{\rm orb}^\prime
   \right) \times \delta \vec{r}_i^{\,\prime},  \label{eq-Vtdt}
\end{equation}
where $\vec{V}_{O_i}^\prime$ is the absolute velocity of the origin of
$O_i X_i Y_i Z_i$ at time $t + \delta t$, $\vec{\Omega}_{\rm
orb}^\prime$ is the orbital angular velocity of the binary at time
$t+\delta t$ in the absence of mass transfer, and $\delta
\vec{\Omega}_{\rm orb}^\prime$ is the perturbation of the orbital
angular velocity at time $t + \delta t$ due to the mass loss/gain of
the binary components.

In the limit of small $\delta t$, taking the difference between the 
linear momenta $\vec{Q}_i^\prime$ and $\vec{Q}_i$, dividing the 
resulting equation by $\delta t$, and noting that the absolute velocity
$\vec{V}_{O_i}$ of the origin of $O_i X_i Y_i Z_i$ at time $t$ is
equal to $\vec{V}_i$, yields
\begin{equation}
M_i \frac{d\vec{V}_{O_i}}{dt} = \vec{F}_i +
  \dot{M}_i\, \vec{U}_{\delta M_i}. \label{eq-dVi}  
\end{equation}
Here, $\vec{F}_i = d\vec{Q}_i/dt$ is the sum of all external forces
acting on star $i$, $\dot{M_i} = dM_i/dt$ is the mass loss/accretion
rate of star $i$, and  
\begin{equation}
\vec{U}_{\delta M_i} = \vec{W}_{\delta M_i} - \vec{V}_{O_i} 
  - \vec{\Omega}_{\rm orb} \times \vec{r}_{A_i}
\label{eq-Ui}
\end{equation}
is the relative velocity of the ejected/accreted mass element with
respect to the ejection/accretion point $A_i$. In the derivation of 
equation~(\ref{eq-Ui}), we have made use of equations (\ref{eq-DelR}) and 
(\ref{eq-Vtdt}) and restricted ourselves to first-order terms in 
the small quantities $\delta M_i$ and $\delta \vec{\Omega}_{\rm 
orb}^\prime$.

The absolute acceleration of the center of mass of star $i$ with
respect to the inertial frame $OXYZ$ is given by 
\begin{equation}
\frac{d^2 \vec{R}_i}{dt^2} = \vec{\gamma}_{O_i}
  + \vec{\gamma}_{\rm rel,i} + \vec{\gamma}_{\rm cor,i},
\end{equation}
where $\vec{\gamma}_{O_i} = d\vec{V}_{O_i}/dt$ is the
acceleration of the origin of $O_i X_i Y_i Z_i$ with respect to
$OXYZ$, $\vec{\gamma}_{\rm rel,i} = (\ddot{M}_i/M_i)\, \vec{r}_{A_i}$
is the relative acceleration of the center of mass of star $i$ with
respect to $O_i$, and $\vec{\gamma}_{\rm cor,i} = 2\,
(\dot{M}_i/M_i)\, (\vec{\Omega}_{\rm orb} \times \vec{r}_{A_i})$ is
the Coriolis acceleration of the center of mass of star $i$ with
respect to $O_i$\footnote{The centrifugal acceleration does not play a
role since it is proportional to $\delta \vec{r}_i^{\,\prime}$ which
vanishes for small $\delta t$.}. The expressions for
$\vec{\gamma}_{\rm rel,i}$ and $\vec{\gamma}_{\rm cor,i}$ follow from
the observation that $d\vec{\rho}^{\,\prime}_{A_i}/dt = 
(\dot{M}_i/M_i)\,
\vec{r}_{A_i}$, which one obtains by dividing equation~(\ref{eq-Delrho}) by
$\delta t$ in the limiting case of small $\delta t$.  The equation of
motion for the mass center of star $i$ with respect to the inertial
frame $OXYZ$ then becomes
\begin{eqnarray}
M_i \frac{d^2 \vec{R}_i}{dt^2} &=& \vec{F}_i 
  + \dot{M}_i \left( \vec{U}_{\delta M_i} + 2\,
  \vec{\Omega}_{\rm orb} \times \vec{r}_{A_i} \right)
  \nonumber \\ 
 &+& \ddot{M}_i\, \vec{r}_{A_i}.
\label{eq-mtn}
\end{eqnarray}

\subsection{Relative Motion of the Binary Components}
\label{relmot}

We can now obtain the equation describing the relative motion of the
accretor (star~2) with respect to the donor (star~1) by taking the
difference of the equations of motion of the stars with respect to
the inertial frame of reference. For convenience, we first decompose
the sum of the external forces acting on each star as
\begin{equation}
\vec{F}_i = - \frac{G\,M_1\,M_2}{\left| \vec{r} \right|^2}\,
  \frac{\vec{R}_i}{| \vec{R}_i |} + \vec{f}_i,
\end{equation}
where $G$ is the Newtonian constant of gravitation, and $\vec{f}_i$
the total gravitational force exerted on star $i$ by the particles in
the mass-transfer stream. It follows that
\begin{eqnarray}
\label{eq-R21}
\frac{d^2\vec{r}}{dt^2} &=& -\frac{G \left( M_1+M_2
    \right)}{\left| \vec{r} \right|^3}\, \vec{r} 
    + \frac{\vec{f}_2}{M_2}
    - \frac{\vec{f}_1}{M_1}  \nonumber \\
&+& \frac{\dot{M}_2}{M_2} \left( \vec{v}_{\delta M_2} +
    \vec{\Omega}_{\rm orb} \times \vec{r}_{A_2} \right) \nonumber \\
&-& \frac{\dot{M}_1}{M_1} \left( \vec{v}_{\delta M_1} +
    \vec{\Omega}_{\rm orb} \times \vec{r_{A_1}} \right) \\
&+& \frac{\ddot{M_2}}{M_2}\, \vec{r}_{A_2} 
  - \frac{\ddot{M_1}}{M_1}\, \vec{r}_{A_1}, \nonumber 
\end{eqnarray}
where $\vec{r} = \vec{R}_2 - \vec{R}_1$ is the position vector of the
accretor with respect to the donor, and $\vec{v}_{\delta M_i} =
\vec{W}_{\delta M_i} - \vec{V}_{i}$ is the velocity of the
ejected/accreted mass element with respect to the mass center of the
mass losing/gaining star. 

Equation~(\ref{eq-R21}) can be written in the form of a perturbed
two-body problem as
\begin{equation}
\frac{d^2\vec{r}}{dt^2} = -\frac{G \left( M_1+M_2 \right)}
  {\left| \vec{r} \right|^3}\,
  \vec{r} + S\, \hat{x} + T\, \hat{y} + W\, \hat{z},
\end{equation}
where $\hat{x}$ is a unit vector in the direction of $\vec{r}$,
$\hat{y}$ is a unit vector in the orbital plane perpendicular to
$\vec{r}$ in the direction of the orbital motion, and $\hat{z}$ is a
unit vector perpendicular to the orbital plane parallel to and in the
same direction as $\vec{\Omega}_{\rm orb}$.  The functions $S$, $T$,
and $W$ are found by taking the dot product of the perturbing force 
arising from the mass transfer between the binary components and the 
unit vector in the $\hat{x}$, $\hat{y}$, and $\hat{z}$ directions, 
respectively.  These vector components are
\begin{eqnarray}
S &=& \frac{f_{2,x}}{M_2} - \frac{f_{1,x}}{M_1}  
+ \frac{\dot{M}_2}{M_2} \left( v_{\delta M_2,x} - |\vec{\Omega}_{\rm orb}|
|\vec{r}_{A_2}| \sin{\phi} \right) \nonumber \\
&-& \frac{\dot{M}_1}{M_1} v_{\delta M_1,x} 
+ \frac{\ddot{M}_2}{M_2} |\vec{r}_{A_2}| \cos{\phi}
- \frac{\ddot{M}_1}{M_1} |\vec{r}_{A_1}|, \label{eq-S} \\
T &=& \frac{f_{2,y}}{M_2} - \frac{f_{1,y}}{M_1}  
+ \frac{\dot{M}_2}{M_2} \left( v_{\delta M_2,y} 
+ |\vec{\Omega}_{\rm orb}| |\vec{r}_{A_2}| \cos{\phi} \right) \nonumber \\ 
&-&\frac{\dot{M}_1}{M_1} \left( v_{\delta M_1,y}
+ |\vec{\Omega}_{\rm orb}||\vec{r}_{A_1}| \right) +
\frac{\ddot{M}_2}{M_2} |\vec{r}_{A_2}| \sin{\phi}, \label{eq-T} \\
W &=& \frac{f_{2,z}}{M_2} - \frac{f_{1,z}}{M_1}, \label{eq-W}
\end{eqnarray}
where $\phi$ is the angle between $\hat{x}$ and the vector from the 
center of mass of the accretor to the mass
accretion point $A_2$, and the subscripts $x$, $y$, and $z$ 
denote vector components in the $\hat{x}$, $\hat{y}$, and $\hat{z}$ 
directions, respectively. In working out the vector products
$\vec{\Omega}_{\rm orb} \times \vec{r}_{A_2}$, we assumed that $A_2$ is 
located on the equator of the accreting star\footnote{For brevity, we refer to the point $A_2$ as lying on the 
stellar surface. Though, in practice, it can lie at any point near the 
star where the 
transferred mass can be considered to be part of the accretor. 
For instance, if an accretion disk has formed around the 
accretor, it would be equally valid to write $A_2$ as the point 
where the transferred mass impacts the outer edge of the accretion 
disk.}.  The terms contributing to the perturbed orbital motion can be 
categorized as follows: (i) term proportional to $\vec{f}_i$ represent 
gravitational perturbation on the binary components caused by mass 
elements in the mass-transfer stream; (ii) terms proportional to 
$\dot{M}_i$ represent linear momentum exchange between the mass donor 
and accretor; and (iii) terms proportional to $\ddot{M}_i$ represent 
shifts in the position of the mass centers of the mass donor and 
accretor due to the non-spherical symmetry of the mass loss or gain.  In 
the limiting case where both stars are treated as point masses
($|\vec{r}_{A_1}| \rightarrow 0$ and $|\vec{r}_{A_2}| \rightarrow 0$), 
the only non-zero terms in the perturbed equations of motion are those 
due to gravitational perturbations of the mass transfer stream and the 
transport of linear momentum.

\subsection{Comparison with Previous Work}
\label{sec-appb}

The most recent study on the orbital evolution of eccentric
mass-transferring binaries has been presented by \citet[][hereafter MW83
and MW84, respectively]{MW83, MW84}. These authors extended the work
of \citet{H56}, \citet{K64-2}, and \citet{P64} by accounting for the
effects of linear momentum transport between the binary components, as
well as possible perturbations to the orbital motion caused by the
mass transfer stream. However, they also derived the equations
describing the motion of the binary components with respect to a frame
of reference with origin at the mass center of the binary, which, for
mass transferring systems, is not an inertial frame of reference. The
equations therefore do not account for the accelerations of the binary
mass center caused by the mass transfer. Here, we
demonstrate that if the procedure adopted by Matese \& Whitmire is
developed with respect to an inertial frame of reference that is not
connected to the binary, the resulting equations are in agreement with
those derived in \S\,\ref{relmot}.

The core of Matese \& Whitmire's derivation is presented in Section II
of MW83. While the authors choose to adopt a reference frame with
origin at the binary mass center early on in the investigation, the
choice of the frame does not affect the derivation of the equations of
motion up to and including their equation~(24). In particular, equation~(13) in
MW83, which, in our notation, reads
\begin{equation}
\label{eq-apb-p}
\vec{p}_i = M_i \dot{\vec{R}}_i - \dot{M}_i\vec{r}_{A_i}, 
\end{equation}
is valid with respect to any inertial frame of reference with
arbitrary position and orientation in space. The same applies to equation (1)
in MW84:
\begin{equation}
\label{eq-apb-pdot}
\dot{\vec{p}}_i = -G M_1 M_2\,
  \frac{\vec{R}_i-\vec{R}_{3-i}}{|\vec{R}_i-\vec{R}_{3-i}|^3}
  + \vec{f}_i + \vec{\Psi}_i.
\end{equation}
In these equations, $\vec{p}_i$ is the linear momentum of star $i$,
and $\vec{\Psi}_i=\dot{M}_i (\dot{\vec{R}}_i+\vec{v}_{\delta M_i} )$
is the amount of linear momentum transported by the transferred mass
per unit time (see equation~(3) of MW84). Substitution of
equation~(\ref{eq-apb-p}) into equation~(\ref{eq-apb-pdot}) then yields
\begin{eqnarray}
M_i\, \ddot{\vec{R}}_i = &-& GM_{3-i}\,
  \frac{\vec{R}_i-\vec{R}_{3-i}}{|\vec{R}_i-\vec{R}_{3-i}|^3} \nonumber \\
  &+& \frac{\vec{f}_i}{M_i} + \frac{\dot{M}_i}{M_i} 
  \left( \vec{v}_{\delta M_i} +
  \dot{\vec{r}}_{A_i} \right) + \frac{\ddot{M}_i}{M_i}\vec{R}_{A_i},
\label{eq-apb-mrddot}
\end{eqnarray}
and thus
\begin{eqnarray}
\label{eq-apb-eqm}
\frac{d^2\vec{r}}{dt^2} = &-&\frac{G \left( M_1+M_2
  \right)}{\left| \vec{r} \right|^3}\, \vec{r}
  + \frac{\vec{f}_2}{M_2} - \frac{\vec{f}_1}{M_1} \nonumber \\
  &+& \frac{\dot{M}_2}{M_2} \left( \vec{v}_{\delta M_2} +
    \dot{\vec{r}}_{A_2} \right)
  - \frac{\dot{M}_1}{M_1} \left( \vec{v}_{\delta M_1} +
    \dot{\vec{r}}_{A_1} \right) \nonumber \\
  &+& \frac{\ddot{M_2}}{M_2}\, \vec{r}_{A_2} 
  - \frac{\ddot{M_1}}{M_1}\, \vec{r}_{A_1}, 
\end{eqnarray}
where $\vec{r} = \vec{R}_2 - \vec{R}_1$.  Setting $\dot{\vec{r}}_{A_i}
= \vec{\Omega}_{\rm orb} \times \vec{r}_{A_i}$, this equation is in
perfect agreement with equation~(\ref{eq-R21}) derived in
\S\,\ref{relmot}.

In MW83 and MW84, the authors incorrectly set $\vec{R}_1=-M_2\,
\vec{r}/(M_1+M_2)$ and $\vec{R}_2=M_1\, \vec{r}/(M_1+M_2)$ in
equation~(\ref{eq-apb-mrddot}), which is valid only when the origin of the
frame of reference coincide with the mass center of the binary. For a
mass-transferring binary, such a frame is, however, not an inertial
frame and can therefore not be used for the derivation of the
equations of motion of the binary components. Instead of
equation~(\ref{eq-apb-eqm}), Matese \& Whitmire therefore find
equations~(7)--(8) in MW84, which lack the terms associated with the
acceleration of the binary mass center due to the mass transfer.

\section{Orbital Evolution Equations}
\label{sec-sec}

\subsection{Secular Variation of the Orbital Elements}

In the classical framework of the theory of osculating elements, the
equations governing the rate of change of the orbital semi-major axis 
$a$ and eccentricity $e$ due to mass transfer are obtained from the 
perturbing functions $S$ and $T$ as \citep[see, e.g.,][]{S60, BC61, D62, 
F70}
\begin{equation}
\frac{da}{dt} = \frac{2}{n(1-e^2)^{1/2}} [ S e \sin{\nu} + T ( 1
  + e \cos{\nu} )],
\label{eq-dadt}
\end{equation}
\begin{eqnarray}
\lefteqn{\frac{de}{dt} = \frac{(1-e^2)^{1/2}}{na}} 
  \nonumber \\
 & & \times \left\{ S \sin{\nu} +
  T \left[ \frac{2\cos{\nu} +e \left(1+\cos^2{\nu}
  \right)}{1+e\cos{\nu}} \right] \right\},
\label{eq-dedt}
\end{eqnarray}
where $n=2\pi/P_{\rm orb}$ is the mean motion and $\nu$ the true
anomaly. These equations are independent of the perturbing function
$W$ which solely appears in the equations governing the rates of
change of the orbital inclination, the longitude of the ascending
node, and the longitude of the periastron. 

After substitution of equations~(\ref{eq-S}) and (\ref{eq-T}) for $S$ and
$T$ into equations~(\ref{eq-dadt}) and~(\ref{eq-dedt}), the equations
governing the rates of change of the semi-major axis and eccentricity
contain periodic as well as secular terms. Here we are mainly
interested in the long-term secular evolution of the orbit, and so we
remove the periodic terms by averaging the equations over one orbital
period:
\begin{equation}
\left< {\frac{da}{dt}} \right>_{\rm sec} \equiv 
  \frac{1}{P_{\rm orb}} \int_{-P_{\rm orb}/2}^{P_{\rm orb}/2}
  {\frac{da}{dt}}\, dt,
\label{eq-dadtsec}
\end{equation}
\begin{equation}
\left< {\frac{de}{dt}} \right>_{\rm sec} \equiv 
  \frac{1}{P_{\rm orb}} \int_{-P_{\rm orb}/2}^{P_{\rm orb}/2}
  {\frac{de}{dt}}\, dt.
\label{eq-dedtsec}
\end{equation}
The integrals in these definitions are most conveniently computed in
terms of the true  anomaly, $\nu$. We therefore make a change of 
variables using
\begin{equation}
dt=\frac{(1-e^2)^{3/2}}{n(1+e\cos{\nu})^2}\,d\nu.
\label{eq-dnudt}
\end{equation}
For binaries with eccentric orbits, the resulting integrals can be
calculated analytically only for very specific functional
prescriptions of the mass-transfer rate $\dot{M}_1$ (e.g., when
$\dot{M}_1$ is approximated by a Dirac delta function centered on the
periastron, see \S\,\ref{sec-delta}). In general, the
integrals must be computed numerically.

\subsection{Conservation of Orbital Angular Momentum}

\label{sec-angmom}

Since the perturbing functions $S$ and $T$ depend on the properties
of the mass transfer stream, calculation of the rates 
of secular
change of the orbital semi-major axis and eccentricity, in principle,
requires the calculation of the trajectories of the particles in the
stream \citep[cf.][]{H69b}. As long as no mass is lost from the
system, such a calculation automatically incorporates the conservation
of total angular momentum in the system. Special cases of angular
momentum conservation can, however, be used to bypass the
calculation of detailed particle trajectories. Here, we adopt such a
special case and assume that any orbital angular momentum carried by
the particles in the mass-transfer stream is always immediately
returned to the orbit, so that the orbital angular momentum of the
binary is conserved.

The orbital angular momentum of a binary with a semi-major axis $a$ and 
eccentricity $e$ is given by
\begin{equation}
J_{\rm orb} = M_1M_2 \left[ \frac{Ga(1-e^2)}{M_1+M_2} \right]^{1/2},
\label{eq-J}
\end{equation}
so that
\begin{equation}
\frac{\dot{J}_{\rm orb}}{J_{\rm orb}} = \frac{\dot{M_1}}{M_1} +
\frac{\dot{M_2}}{M_2}  
- \frac{1}{2}\frac{\dot{M_1}+\dot{M_2}}{M_1+M_2}
+ \frac{1}{2}\frac{\dot{a}}{a} - \frac{e\,\dot{e}}{1-e^2},
\label{eq-Jdot}
\end{equation}
where a dot indicates the time derivative.

In the case of eccentric orbits, substitution of 
equations~(\ref{eq-dadt}) and (\ref{eq-dedt}) into equation 
(\ref{eq-Jdot}) leads to 
\begin{eqnarray}
\frac{\dot{J}_{\rm orb}}{J_{\rm orb}} &=& 
  \frac{\dot{M_1}}{M_1} + \frac{\dot{M_2}}{M_2} 
  - \frac{1}{2}\frac{\dot{M_1}+\dot{M_2}}{M_1+M_2} \nonumber \\ 
&+& \frac{(1-e^2)^{1/2}}{n\,a \left(1+e\cos{\nu} \right)}\,T.
\label{eq-JT}
\end{eqnarray}
As we shall see in the next section, by setting $\dot{M}_1 + \dot{M}_2 =
0$ and $< \dot{J}_{\rm orb}/J_{\rm orb} >_{\rm sec} =0$ and substituting
equation~(\ref{eq-T}) for $T$, equation~(\ref{eq-JT}) allows us to
calculate the $\hat{y}$-component of the final velocities of the 
accreting
particles as a function of their initial velocities without resorting to
the computation of the ballistic trajectories of the mass transfer
stream.  

In the limiting case of a circular orbit, equation~(\ref{eq-Jdot}) is 
usually used to derive the rate of change of the semi-major axis of 
circular binaries under the assumption of conservation of both total 
mass ($\dot{M}_1=-\dot{M}_2$) and orbital angular momentum ($\dot{J}_{\rm 
orb}=0$):
\begin{equation}
\frac{da}{dt} = 2a(\frac{M_1}{M_2}-1)\frac{\dot{M}_1}{M_1}.
\label{circ-da}
\end{equation}

The assumption of orbital angular momentum conservation over secular
timescales ($< \dot{J}_{\rm orb}/J_{\rm orb} >_{\rm sec} =0$) is a
standard assumption in nearly all investigations of conservative mass
transfer in binary systems \citep[e.g., ][]{SPH97, P98}, which is valid
over long timescales provided there is no significant storage of angular
momentum in the spins of the components stars, the accretion flow,
and/or the accretion disk.  In future work, we will investigate the
consequences of both mass and orbital angular momentum losses from the
binary on the evolution of the orbital elements.

\section{Orbital Evolution Timescales}
\label{orbevtim}

In order to assess the timescales of orbital evolution due to mass
transfer in eccentric binaries, we observe that, for eccentric binaries, 
mass transfer is expected to occur first at the periastron of the 
relative orbit, where the component stars are closest to each other.  We 
therefore explore the order of magnitude of the timescales assuming a 
delta function mass transfer profile centered at the periastron of the 
binary orbit
\begin{equation}
\dot{M}_1 = \dot{M}_0\, \delta \left( \nu \right),
\label{eq-del}
\end{equation}
where $\dot{M}_0 < 0 $ is the instantaneous mass transfer rate, and 
$\delta(\nu)$ is the Dirac delta function.

We calculate the rates of secular change of the orbital semi-major axis
and eccentricity from equations~(\ref{eq-dadt})--(\ref{eq-dnudt}), and 
neglect any gravitational attractions exerted by the particles in the
mass-transfer stream on the component stars. Hence, we set
\begin{eqnarray}
f_{1,x} &=& f_{2,x} = 0, \label{eq-f12x} \\
f_{1,y} &=& f_{2,y} = 0. \label{eq-f12y}
\end{eqnarray}

\label{sec-delta}

Substituting equations~(\ref{eq-T}), (\ref{eq-dnudt}), and 
(\ref{eq-del}) -- (\ref{eq-f12y}) into equation~(\ref{eq-JT}) for $< 
\dot{J}_{\rm orb}/J_{\rm orb} >_{\rm sec}=0$ then yields a relationship 
between the initial and final $\hat{y}$-component of the velocities of 
the transferred mass and the initial and final positions of the 
transferred mass given by
\begin{eqnarray}
qv_{\delta M_2,y} &+& v_{\delta M_1,y} = 
 na(1-q)\left(\frac{1+e}{1-e}\right)^{1/2}
- |\vec{\Omega}_{{\rm orb}, P}||\vec{r}_{A_1,P}| \nonumber \\
&-& q|\vec{\Omega}_{{\rm orb}, P}||\vec{r}_{A_2}|\cos{\phi_P} 
\left( \left.1- \frac{d\phi}{d\nu}\right|_{\nu=0} \right),
\label{eq-vr}
\end{eqnarray}
where the subscript $P$ indicates quantities evaluated at the periastron
of the binary orbit, $q=M_1/M_2$ is the binary mass ration, and we have 
used the relation
$d\delta(\nu)/d\nu = -\delta(\nu)/\nu$.  Assuming the transferred mass
elements are ejected by star~1 at the $L_1$ point with a velocity
$\vec{v}_{\delta M_1}$ equal to the star's rotational velocity at $L_1$,
we write
\begin{eqnarray}
\label{eq-vdm1x}
v_{\delta M_1, x} = 0,\\
\label{eq-vdm1y}
v_{\delta M_1,y}=-|\vec{\Omega}_{{\rm orb},P}||\vec{r}_{A_1, P}|.
\end{eqnarray}
Moreover, under the assumption that each periastron passage of the
binary components give rise to an extremum of $\phi(\nu)$, the
derivative $d\phi/d\nu|_{\nu=0}$ is equal to zero in
equation~(\ref{eq-vr}), so that
\begin{equation}
v_{\delta M_2, y} = |\vec{\Omega}_{{\rm orb},P}|\left[ |\vec{r}_{A_1,P}| 
\frac{(1-q)}{q} + |\vec{r}_{A_2}|\cos{\phi_P} \right].
\label{eq-vrel}
\end{equation}

For a binary with orbital period $P_{\rm orb} = 1\,{\rm day}$,
eccentricity $e=0.2$, and component masses $M_1=2\,M_\sun$ and
$M_2=1.44\,M_\sun$, with $|\vec{r}_{A_1,P}|$ the distance from star~1 to
$L_1$ (See Appendix A), and $|\vec{r}_{A_2}|\cos{\phi_P}\approx
2.9\times10^5\,{\rm km}$ (the circularization radius around a compact
object for these binary parameters; see \citet{FKR}), the accreting
matter has a $\hat{y}$-velocity component of the order of $\sim-45\,{\rm
km}\,{\rm s}^{-1}$.

After substitution of equation~(\ref{eq-dnudt}) and 
equations~(\ref{eq-f12x})--(\ref{eq-vr}), the integrals in
equations~(\ref{eq-dadtsec}) and (\ref{eq-dedtsec}) for the rates of
secular change of the orbital semi-major axis and eccentricity can be
solved analytically to obtain
\begin{eqnarray}
\left< \frac{da}{dt} \right>_{\rm sec} &=& 
\frac{a}{\pi}\frac{\dot{M}_0}{M_1}\frac{1}{(1-e^2)^{1/2}}\left[
qe\frac{|\vec{r}_{A_2}|}{a}\cos{\phi_P} \right. \nonumber \\
 &+& \left. e\frac{|\vec{r}_{A_1,P}|}{a} +
(q-1)(1-e^2) \right],
\label{eq-deltaa} 
\end{eqnarray}
\begin{eqnarray}
\left< \frac{de}{dt} \right>_{\rm sec} &=& 
\frac{(1-e^2)^{1/2}}{2\pi} \frac{\dot{M}_0}{M_1}\left[ 
q\frac{|\vec{r}_{A_2}|}{a}\cos{\phi_P} \right. \nonumber \\
&+& \left. \frac{|\vec{r}_{A_1,P}|}{a} + 2(1-e)(q-1) \right].
\label{eq-deltae}
\end{eqnarray}
We note that for a delta-function mass transfer
rate given by equation~(\ref{eq-del}) the $\hat{x}$-component of the
velocities $\vec{v}_{\delta M_1}$ and $\vec{v}_{\delta M_2}$ does not
enter into the derivation of theses equations due to the $\sin{\nu}$ 
term in equations~(\ref{eq-dadt}) and (\ref{eq-dedt}).  Furthermore, in 
the limiting case of a circular orbit,
equation~(\ref{eq-deltaa}) reduces to equation~(\ref{circ-da}), provided that
$\dot{M_1}$ in that equation is interpreted as the secular mean mass
transfer rate $\left<\right.\!\dot{M_1}\!\left.\right>_{\rm
sec}=\dot{M}_0/(2\pi)$.  In Appendix~B, we present an alternative
derivations to equations~(\ref{eq-deltaa}) and (\ref{eq-deltae}) in the
limiting case where the stars are treated as point masses.

The rates of secular change of the semi-major axis and orbital
eccentricity are thus linearly proportional to the magnitude of the
mass transfer rate at periastron. Besides the obvious dependencies on
$a$, $e$, $q$, and $M_1$, the rates also depend on the ratio of the
donor's rotational angular velocity $\Omega_1$ to the orbital angular
velocity $\Omega_{{\rm orb},P}$ at periastron through the position of 
the $L_1$
point, $\vec{r}_{A_1}$. A fitting formula for the position of the
$L_1$ point accurate to better than $4\,\%$ over a wide range of $q$,
$e$, and $\Omega_1/\Omega_{{\rm orb},P}$ is given by equation~(\ref{XL1fit}) in
Appendix~\ref{sec-appa}. While the fitting formula can be used to
obtain fully analytical rates of secular change of the semi-major axis
and eccentricity, we here use the exact solutions for the
position of $L_1$ obtained by numerically solving equation~(\ref{XL}) in
Appendix~\ref{sec-appa}.  For a detailed discussion of the 
properties of the $L_1$ point in eccentric binaries, we refer the 
interested reader to \citet{SWK07}.

\begin{figure*}
\plottwo{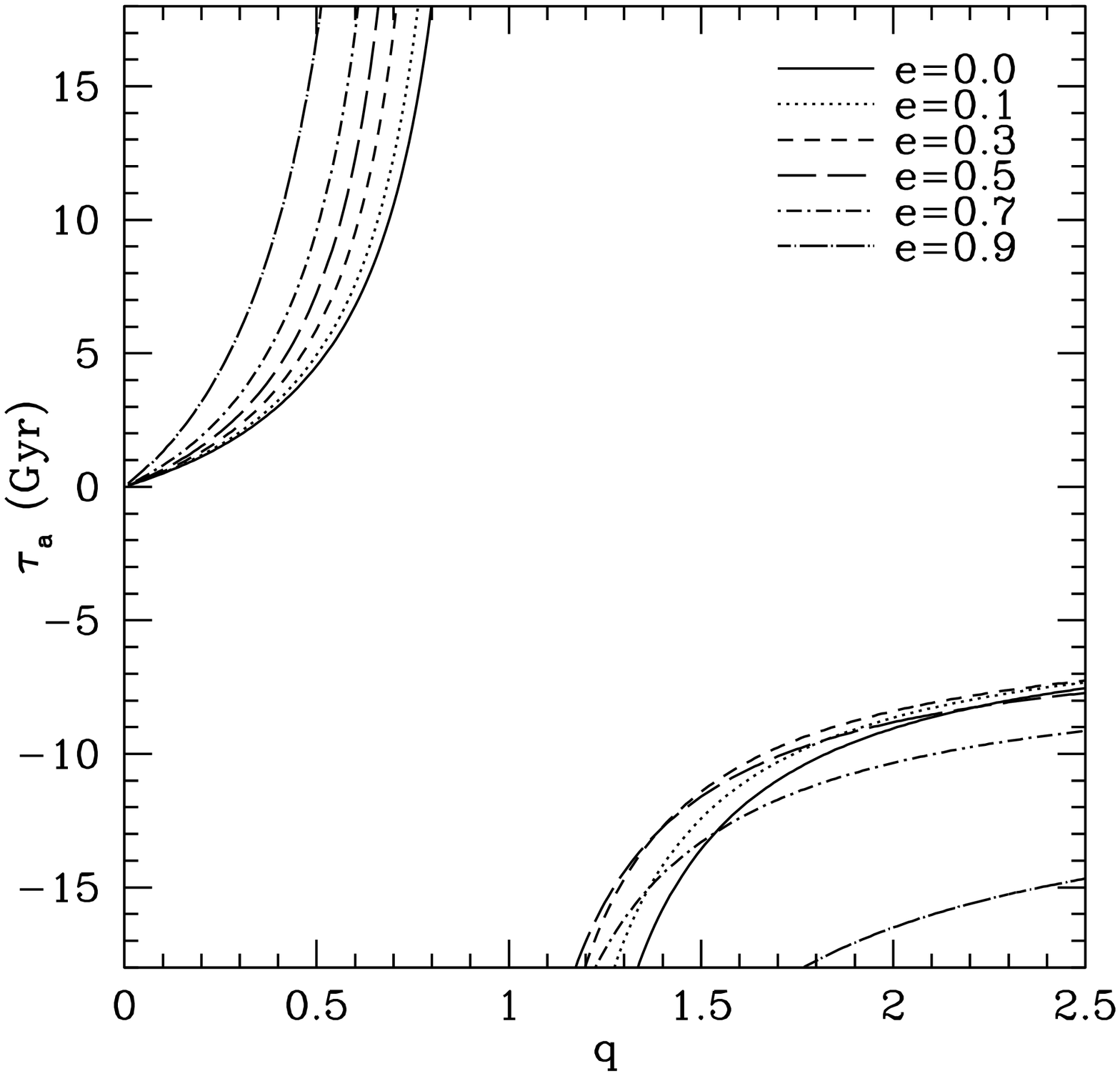}{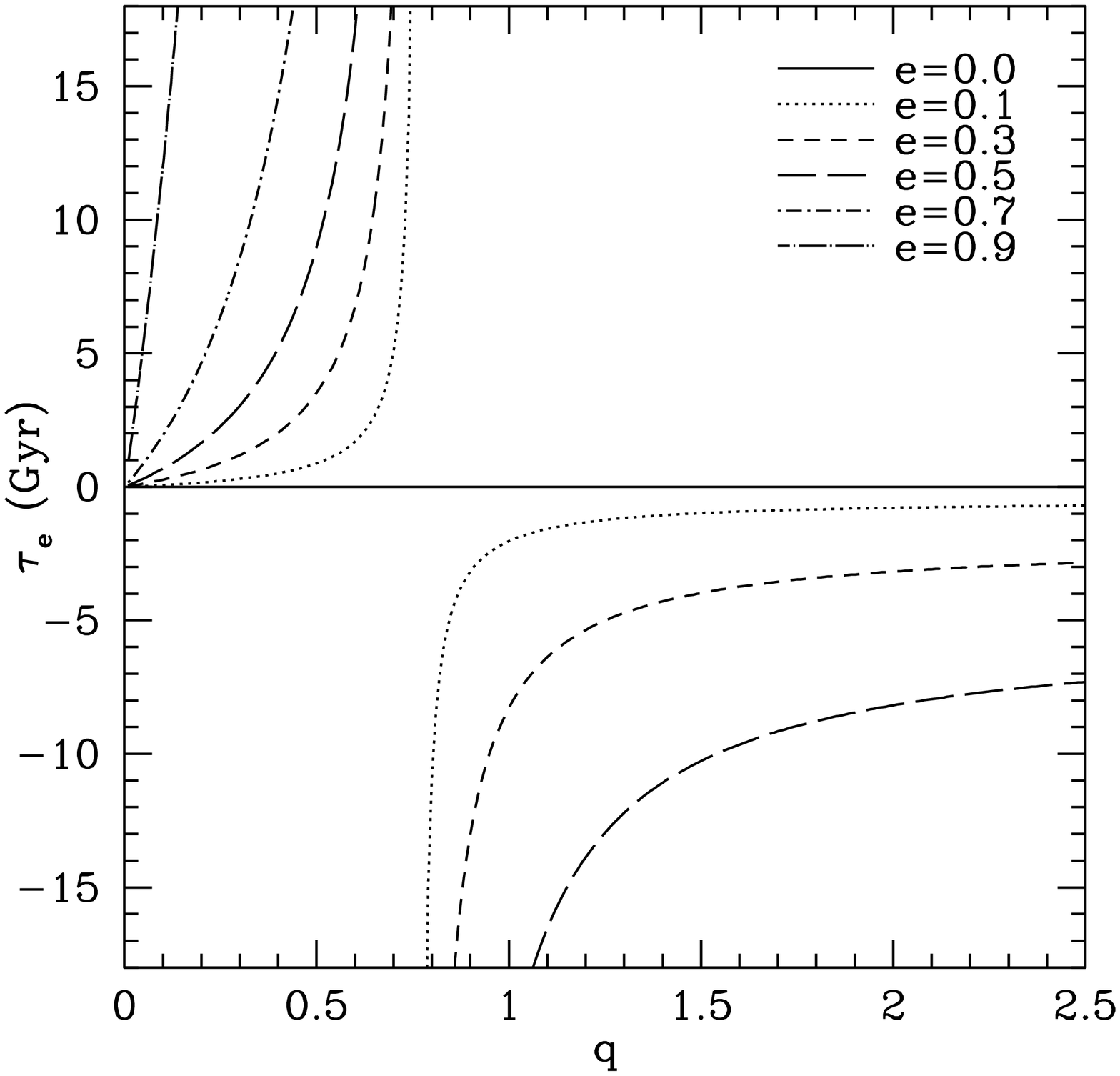}
\caption{Orbital evolution timescales for a delta function mass
  transfer profile centered at the periastron of the binary orbit with 
  an instantaneous mass transfer rate of $\dot{M}_0 = - 
  10^{-9}\,M_\sun\,{\rm yr}^{-1}$. The timescales are calculated under
  the assumption that the donor rotates synchronously with the orbital
  angular velocity at periastron, and that the accretor is a
  $1.44\,M_\odot$ neutron star.  Shown at left (right) are the
  timescales for the evolution of the semi-major axis (eccentricity) as
  a function of the mass ratio, $q$, for a range of eccentricities, $e$.
  Regimes where the timescale is negative correspond to a decrease of
  the semi-major axis (eccentricity), while regimes where the
  timescale is positive correspond to an increase of the semi-major
  axis (eccentricity).
}
\label{delta1}
\end{figure*}

To explore the effects of mass transfer on the orbital elements of
eccentric binaries, we calculate the rates of secular change of the
semi-major axis and eccentricity and determine the characteristic
timescales $\tau_a = a/\dot{a}$ and $\tau_e = e/\dot{e}$. While the
actual timescales are given by the absolute values of $\tau_a$
and $\tau_e$, we here allow the timescales to be negative as well as
positive in order to distinguish negative from positive rates of
secular change of the orbital elements. We also note that since
$|\vec{r}_{A_1, P}| \propto a$ (see Appendix A), the timescales do not
explicitly depend on the orbital semi-major axis $a$ except through 
the ratio $|\vec{r}_{A_2}|/a$ of the radius of the accretor to the 
semi-major axis.  For convenience, we therefore assume the accretor to 
be a compact object with radius $|\vec{r}_{A_2}| << a$.  The timescales 
are found to be insensitive to terms containing $|\vec{r}_{A_2}|/a$ in 
equation (\ref{eq-deltaa}) and (\ref{eq-deltae}).  Varying 
$|\vec{r}_{A_2}|$ from $0$ to $0.01a$ changes the timescales by less 
that 10\%.  In what follows, we therefore set $|\vec{r}_{A_2}|=0$.  An
implicit dependence on $a$ may then still occur through the amplitude 
$\dot{M}_0$ of the mass transfer rate at periastron. Since incorporating 
such a dependence in the analysis requires detailed modeling of the 
evolution of the donor star, which is beyond the scope of this 
investigation, we here restrict ourselves to exploring the timescales of 
orbital evolution for a constant $\dot{M}_0$. The linear dependence of 
$\left <\dot{a} \right >_{\rm sec}$ and $\left < \dot{e} \right >_{\rm 
sec}$ on $\dot{M}_0$ in any case allows for any easy rescaling of 
our results to different mass transfer rates.  

In Fig.~\ref{delta1}, we show the variations of $\tau_a$ and $\tau_e$ as
functions of $q$ for $\dot{M}_0 = -10^{-9}\,M_\sun\,{\rm yr^{-1}}$ and
$e=0.0,0.1, \ldots, 0.9$. In all cases, the donor is assumed to rotate
synchronously with the orbital angular velocity at the periastron, and
the accretor is assumed to be a neutron star of mass
$M_2=1.44\,M_\odot$.  The timescales of the secular evolution of the
semi-major axis show a strong dependence on $q$, and a milder dependence
on $e$, unless $e \ga 0.7$. The timescales for the secular evolution of
the orbital eccentricity always depend strongly on both $q$ and $e$. 
These timescales can furthermore be positive as well as negative, so
that the semi-major axis and eccentricity can increase as well as
decrease under the influence of mass transfer at the periastron of the 
binary orbit.

\begin{figure*}
\plottwo{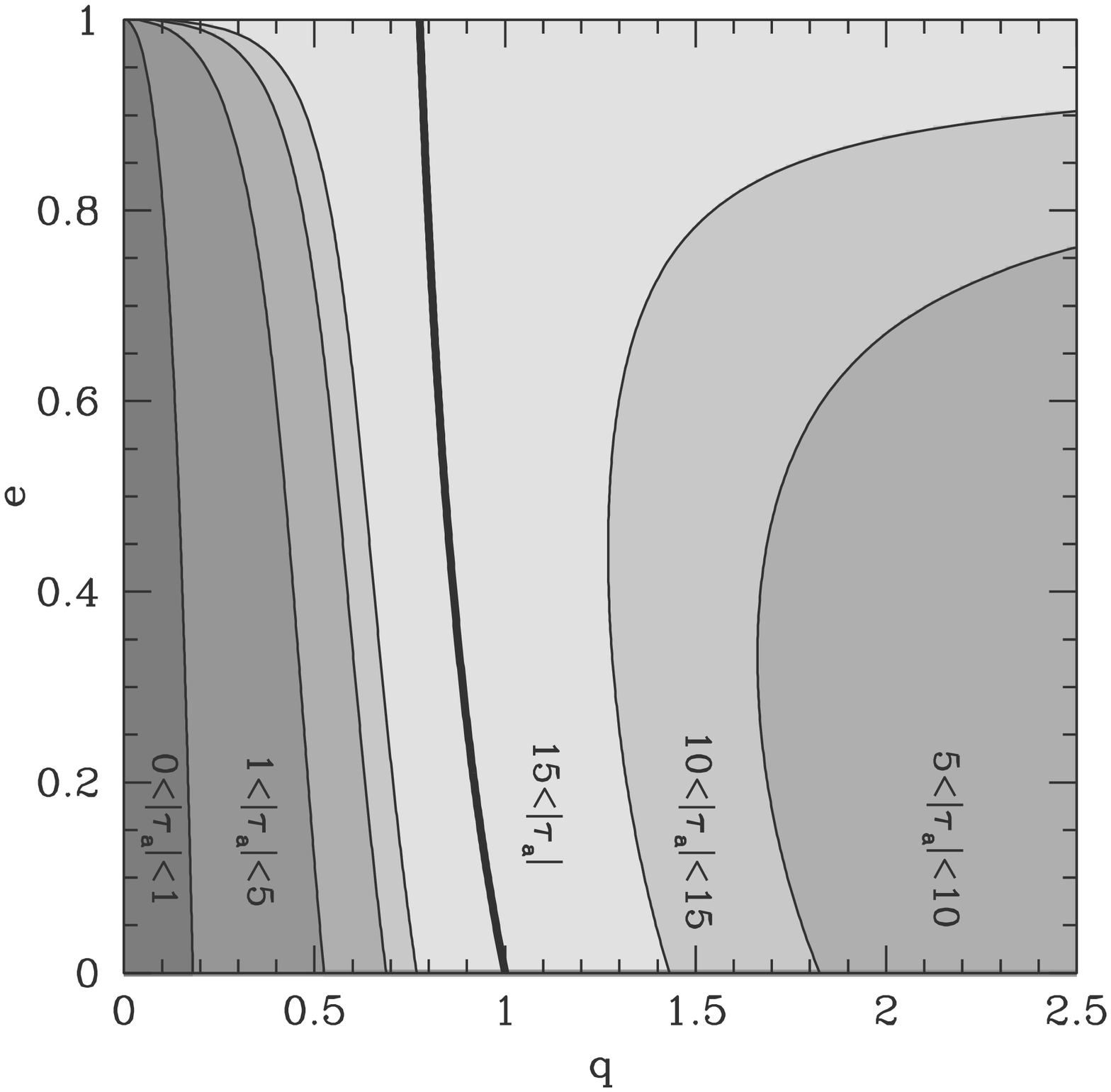}{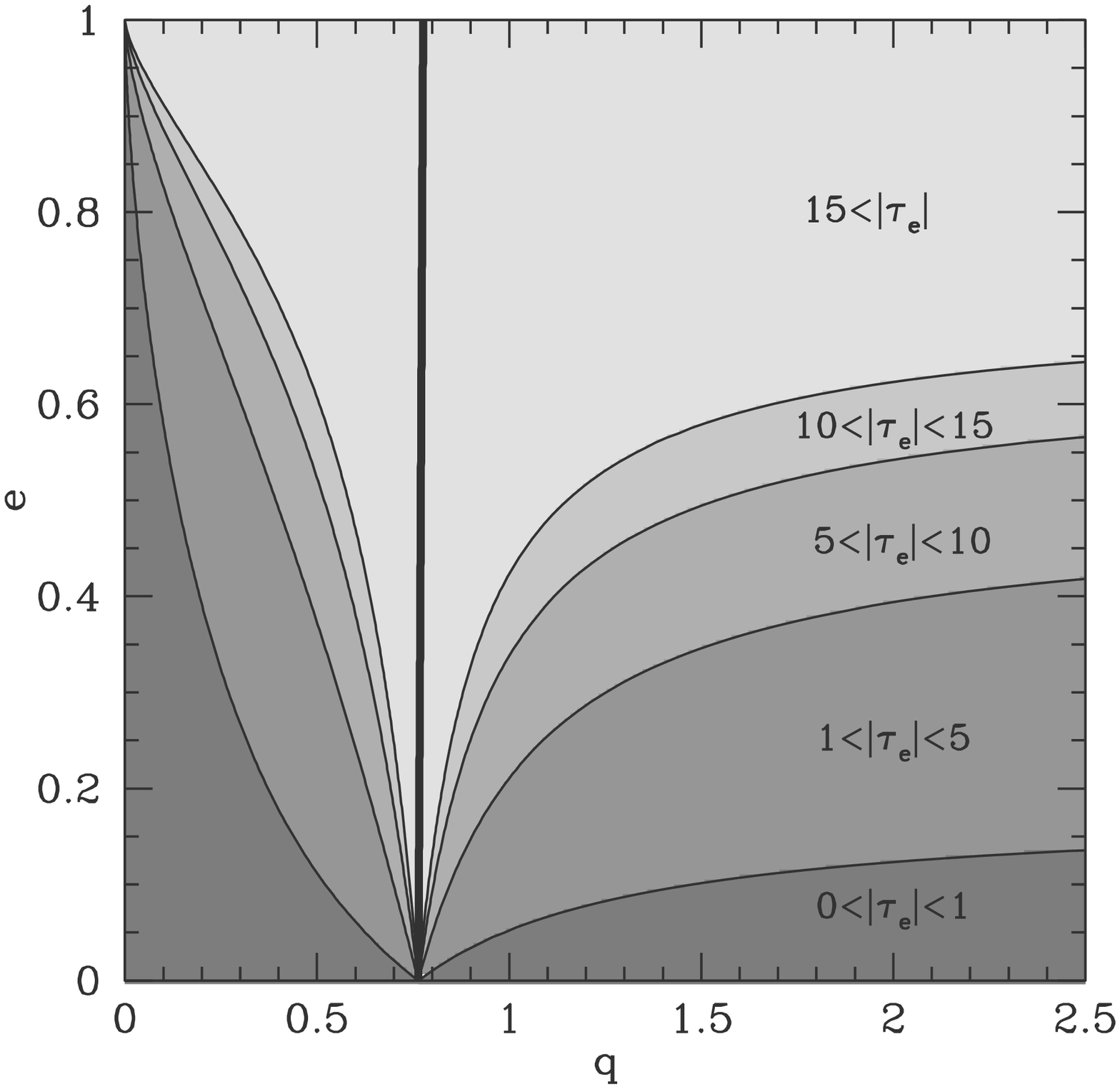}
\caption{Contour plot of the orbital evolution timescales for the 
  semi-major axis and eccentricity in the
  $(q,e)$-plane for the same set of assumptions as adopted in 
  figure~\ref{delta1}.  Timescales for the evolution of the 
  semi-major axis are shown
  on the left; timescales for the evolution of the orbital
  eccentricity on the right. The different shades of gray designate
  regions of the $(q,e)$ parameter space with timescales (in Gyr) in the 
  ranges labeled in the plots.  The thick black line near the center
  of each plot designates the transition point where the rate of
  change of the semi-major axis or orbital eccentricity changes from
  positive (to the left of the thick black line) to negative (to 
  the right of the thick black line).}
\label{contours}
\end{figure*}

From Fig.~\ref{delta1}, as well as equations~(\ref{eq-deltaa}) and
(\ref{eq-deltae}), it can be seen that, for a given ratio of the donor's
rotational angular velocity to the orbital angular velocity at
periastron, the line dividing positive from negative rates of secular
change of the orbital elements is a function of $q$ and $e$. This is
illustrated further in Fig.~\ref{contours} where the timescales of
orbital evolution are displayed as contour plots in the $(q,e)$-plane.
The thick black line near the center of the plots marks the transition
values of $q$ and $e$ where the rates of secular change of $a$ and $e$
transition from being positive (to the left of the thick black line) to
negative (to the right of the thick black line). Varying
$\Omega_1/\Omega_{{\rm orb},P}$ between $0.5$ and $1.5$ changes the
position of the transition line by less than $10\%$ in comparison to the
$\Omega_1/\Omega_{{\rm orb},P}=1$ case displayed in Figures~\ref{delta1}
and \ref{contours}.

In the limiting case of a circular orbit, the orbit expands when
$q<1$ and shrinks when $q>1$, in agreement with the classical result
obtained from equation~(\ref{circ-da}). For non-zero eccentricities, the
critical mass ratio separating positive from negative values of $\left
< \dot{a} \right >_{\rm sec}$ decreases with increasing orbital
eccentricities. This behavior can be understood by substituting the
fitting formula for the position of the $L_1$ point given by
equation~(\ref{XL1fit}) in Appendix~\ref{appL1} into equation~(\ref{eq-deltaa})
and setting $\left< \dot{a} \right>_{\rm sec} =0$.  However, we can fit 
the critical mass ratio separating expanding from shrinking orbits with 
a simpler formula given by
\begin{equation}
q_{\rm crit} \simeq 1 - 0.4e + 0.18e^2.
\label{eq-dadttran}
\end{equation}

The critical mass ratio separating positive from negative values of
$\left < \dot{e} \right >_{\rm sec}$ is largely independent of
$e$. Proceeding in a similar fashion as for the derivation of
equation~(\ref{eq-dadttran}), we derive the critical mass ratio separating
increasing from decreasing eccentricities to be approximately given by
\begin{equation}
q_{\rm crit} \simeq 0.76 + 0.012e.
\label{eq-dedttran}
\end{equation}

Last, we note that a more quantitative numerical comparison between
the above approximation formulae for $q_{\rm crit}$ and the exact
numerical solutions shows that equations~(\ref{eq-dadttran}) and
(\ref{eq-dedttran}) are accurate to better than 1\%.

\section{Tidal Evolution Timescales}

\begin{figure*}
\plottwo{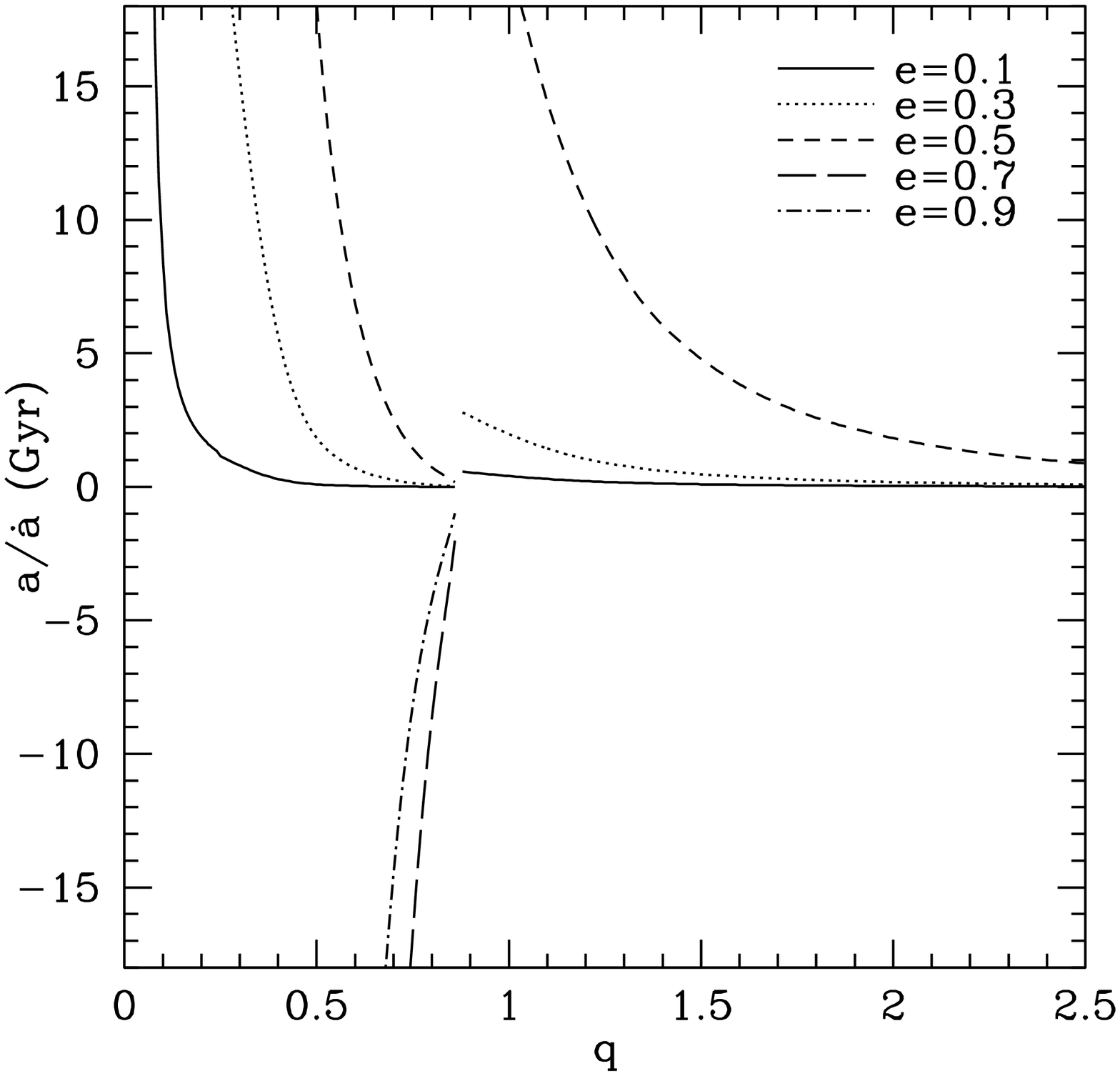}{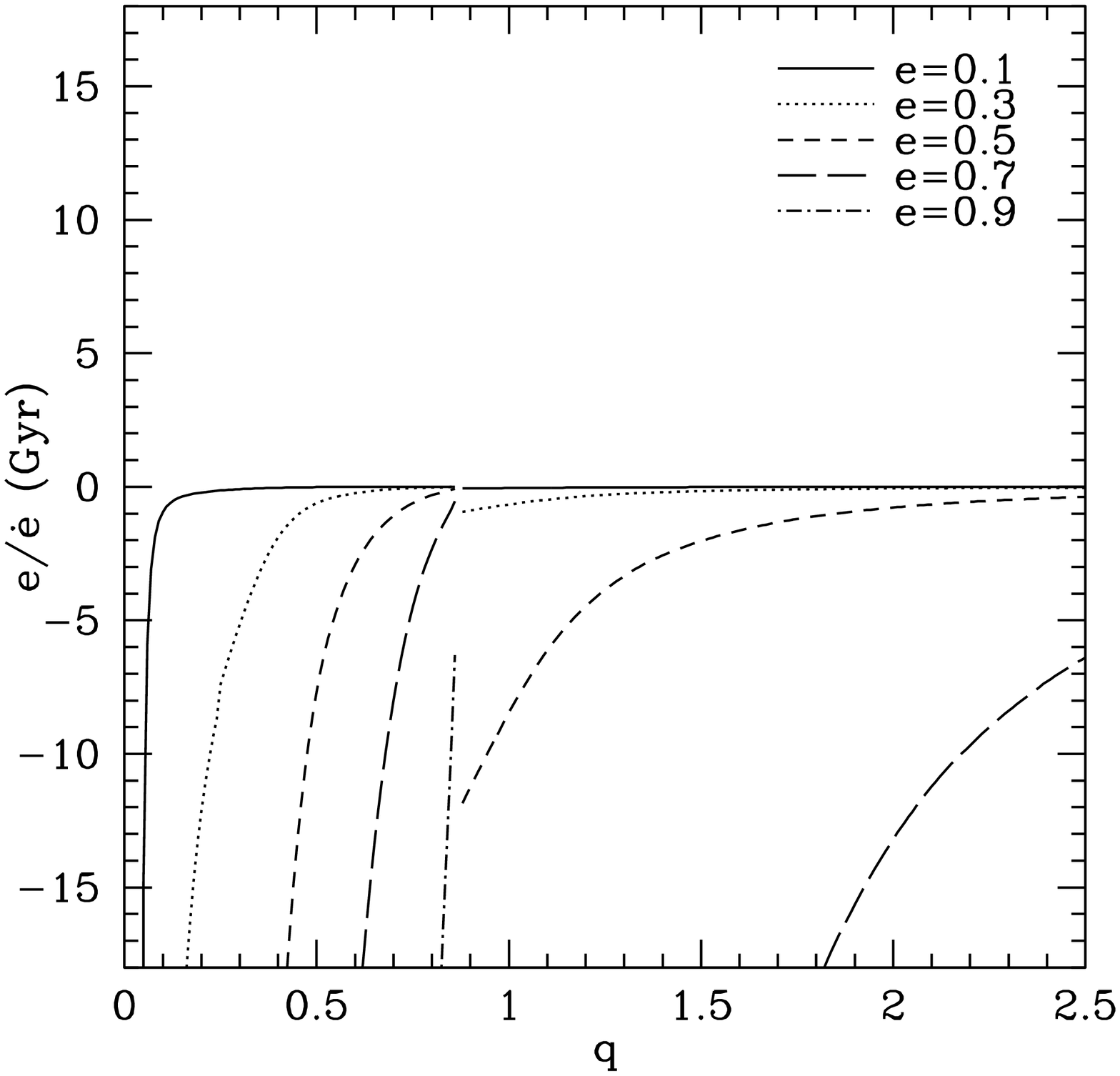}
\caption{Timescales of orbital evolution due to tidal dissipation in a
  Roche-lobe filling component of a close binary under the assumption
  that the donor is a zero-age main-sequence star rotating
  synchronously with the orbital angular velocity at the periastron,
  and the accretor is a $1.44\,M_\odot$ neutron star.  Shown at left
  (right) are the timescales for the evolution of the semi-major axis
  (eccentricity) as a function of the mass ratio, $q$, for 
  different orbital eccentricities $e$.  Regimes
  where the timescales are negative correspond to a decrease of the
  semi-major axis (eccentricity), while regimes where the timescales
  are positive correspond to an increase of the semi-major axis
  (eccentricity). The discontinuity at $q \simeq 0.87$ corresponds to
  the transition from donor stars with convective envelopes ($M_1 \la
  1.25\,M_\odot$) to donor stars with radiative envelopes ($M_1 \ga
  1.25\,M_\odot$).}
\label{fig-tid-lin}
\end{figure*}

A crucial question for assessing the relevance of the work
presented here is how the derived orbital evolution timescales compare
to the corresponding timescales associated with other orbital
evolution mechanisms such as tides. In Fig.~\ref{fig-tid-lin}, we show
the secular evolution timescales of the semi-major axis and orbital
eccentricity of a mass-transferring binary due to tidal dissipation in
the donor star as a function of $q$, for different values of the 
eccentricity, $e$. The timescales are strong functions of 
$|\vec{r}_{A_1}|/a$ and are determined as in \citet{HTP02}\footnote{Note 
that there is a typo in equation~(42) of \citet{HTP02}. The correct equation 
for $k/T$ for stars with radiative envelopes is (J. Hurley, Private 
Communication)
\[
\left( k/T\right)_r =
  1.9782 \times 10^4\left( MR^2/a^5 \right)^{1/2}
  \left(1+q_2\right)^{5/6}E_2\,{\rm yr}^{-1}. \nonumber
\]}
\citep[see also][]{Z77,Z78,H81}.  The radius $|\vec{r}_{A_1}|$ is
determined by assuming the donor is on the zero-age
main sequence and that the orbital separation is then obtained by 
equating the radius of the donor \citep[given by][]{TEA96} to the 
volume-equivalent radius of its Roche lobe at the periastron of the 
binary orbit \citep[see][]{SWK07}.  As before, we assume the donor
rotates synchronously with the orbital motion at periastron and that the 
accretor is a $1.44\,{\rm M}_\sun$ neutron star.

\begin{figure*}
\plottwo{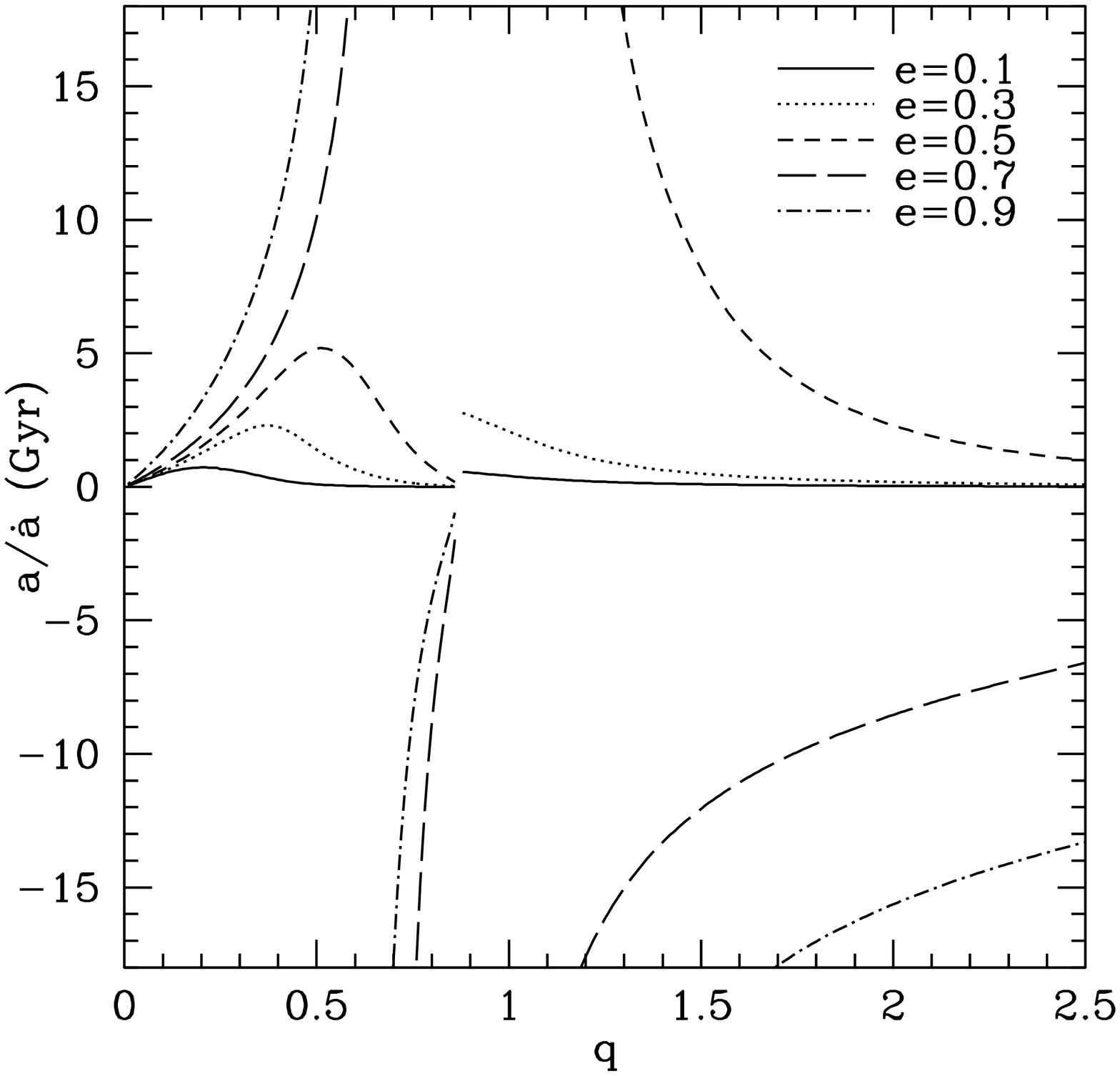}{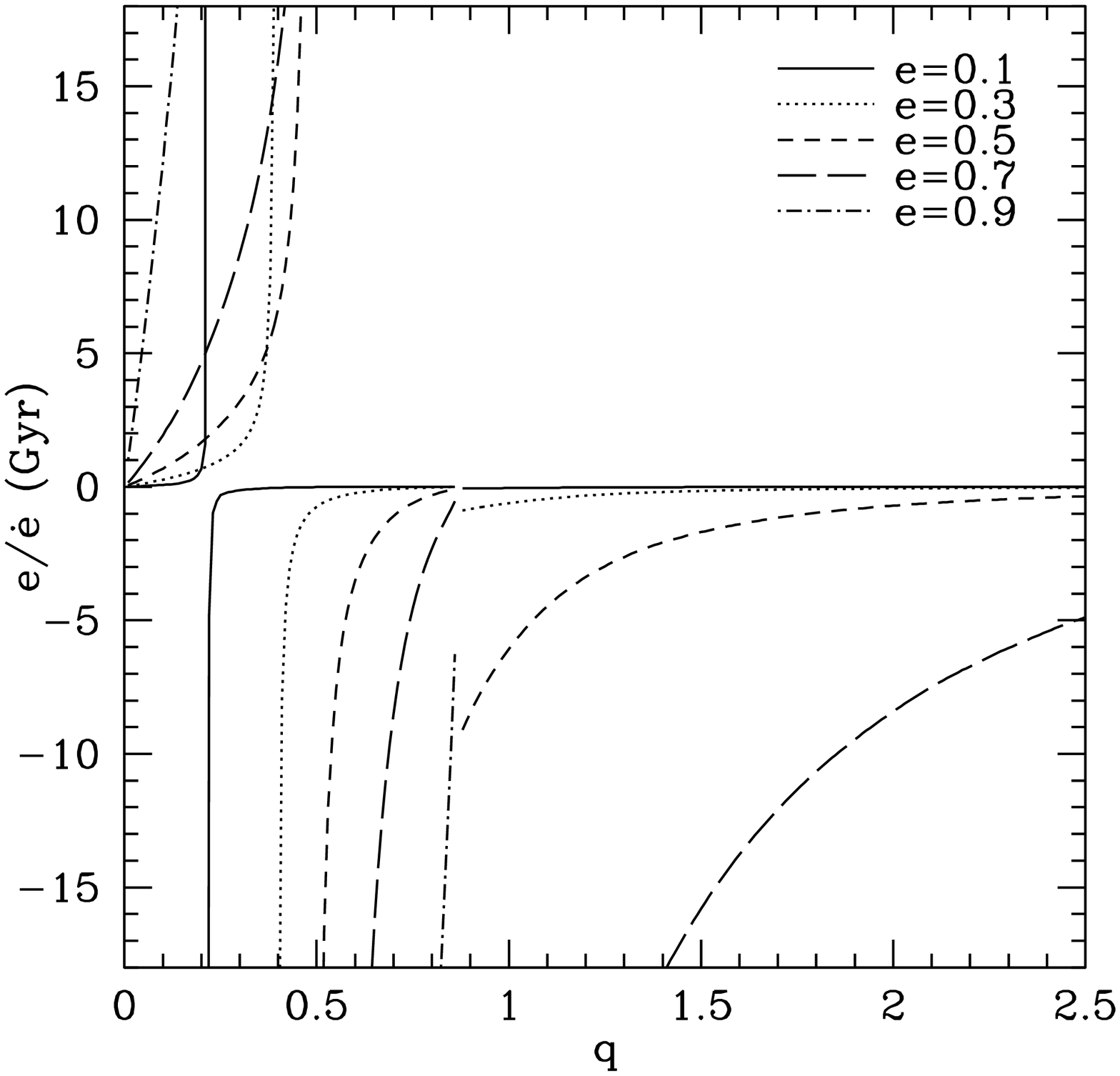}
\caption{Orbital evolution timescales due to the combined effects of 
tidal dissipation in a Roche Lobe filling component of a close binary 
system and a delta function mass transfer with an amplitude 
$\dot{M}_0=-10^{-9}\,M_\sun\,{\rm yr}^{-2}$ centered at the 
periastron of the orbit under the assumption that 
the donor is rotating synchronously with 
the orbital angular velocity at periastron, and the accretor is a 
$1.44\,M_\sun$ neutron star.  The contribution to the orbital 
evolution timescales due to tides is determined under the 
assumption that the donor is zero-age main-sequence star.  Shown at left 
(right) are the timescales 
for the evolution of the semi-major axis (eccentricity) as a function 
of the mass ratio, $q$, for a range of eccentricities, $e$.  Regimes 
where 
the timescale is negative correspond to a decrease in the semi-major 
axis (eccentricity), while regimes where the timescales is positive 
correspond to an increase in the semi-major axis (eccentricity).  The 
discontinuity at $q\simeq0.87$ corresponds to a transition of the 
dominant tidal dissipation mechanism which is different for donor 
stars with convective envelopes ($M_1\lesssim1.25\,M_\sun$) than for 
donor stars with radiative envelopes ($M_1 \gtrsim1.25\,M_\sun$).}
\label{fig-tot_timescale}
\end{figure*}

The timescales of orbital evolution due to tides range from a few Myr to
more than a Hubble time, depending on the binary mass ratio and the
orbital eccentricity. The discontinuity in the timescales at $q \simeq
0.87$ corresponds to the transition from donor stars with convective 
envelopes ($M_1 \la 1.25\,M_\odot$) to donor stars with radiative 
envelopes ($M_1  \ga 1.25\,M_\odot$) which are subject to different 
tidal dissipation mechanisms.  It follows that tides do not 
necessarily lead to rapid circularization during the early stages of 
mass transfer, especially for orbital eccentricities $e \ga 0.3$. 
Furthermore, for the adopted system parameters, the orbital eccentricity 
always decreases, while the orbital semi-major axis can either increase 
or decrease. Hence, in some regions of the parameter space, the effects 
of tides and mass transfer are additive, while in other regions they are
competitive.  This is illustrated in more detail in 
figure~\ref{fig-tot_timescale} where we show the orbital evolution 
timescales due to the combined effect of tides and mass transfer.  In 
the calculations of the timescales, we have assumed that, at the lowest 
order of approximation, the effects of tides and mass transfer are 
decoupled.  The total rate of change of the orbital elements is then
given by the sum of the rate of change of the orbital elements due to 
tides and mass transfer.

When $q \la 0.87$ and $e \ga 0.4$, the effects of tides and mass 
transfer on the orbital semi-major axis are always opposed, with the 
orbital expansion due to mass transfer dominating the orbital shrinkage
due to tides.  In the case of the orbital eccentricity, the 
increase of the eccentricity due to mass transfer dominates the 
decrease due to tides for mass ratios smaller than some critical 
mass ratio which depends strongly on the orbital eccentricity.  Since 
the timescales of orbital evolution
due to mass transfer are inversely proportional to the magnitude
$\dot{M}_0$ of the mass-transfer rate at periastron, the parameter space
where and the extent to which mass transfer dominates increases with the
rate of mass transfer at periastron.

\begin{figure*}
\plottwo{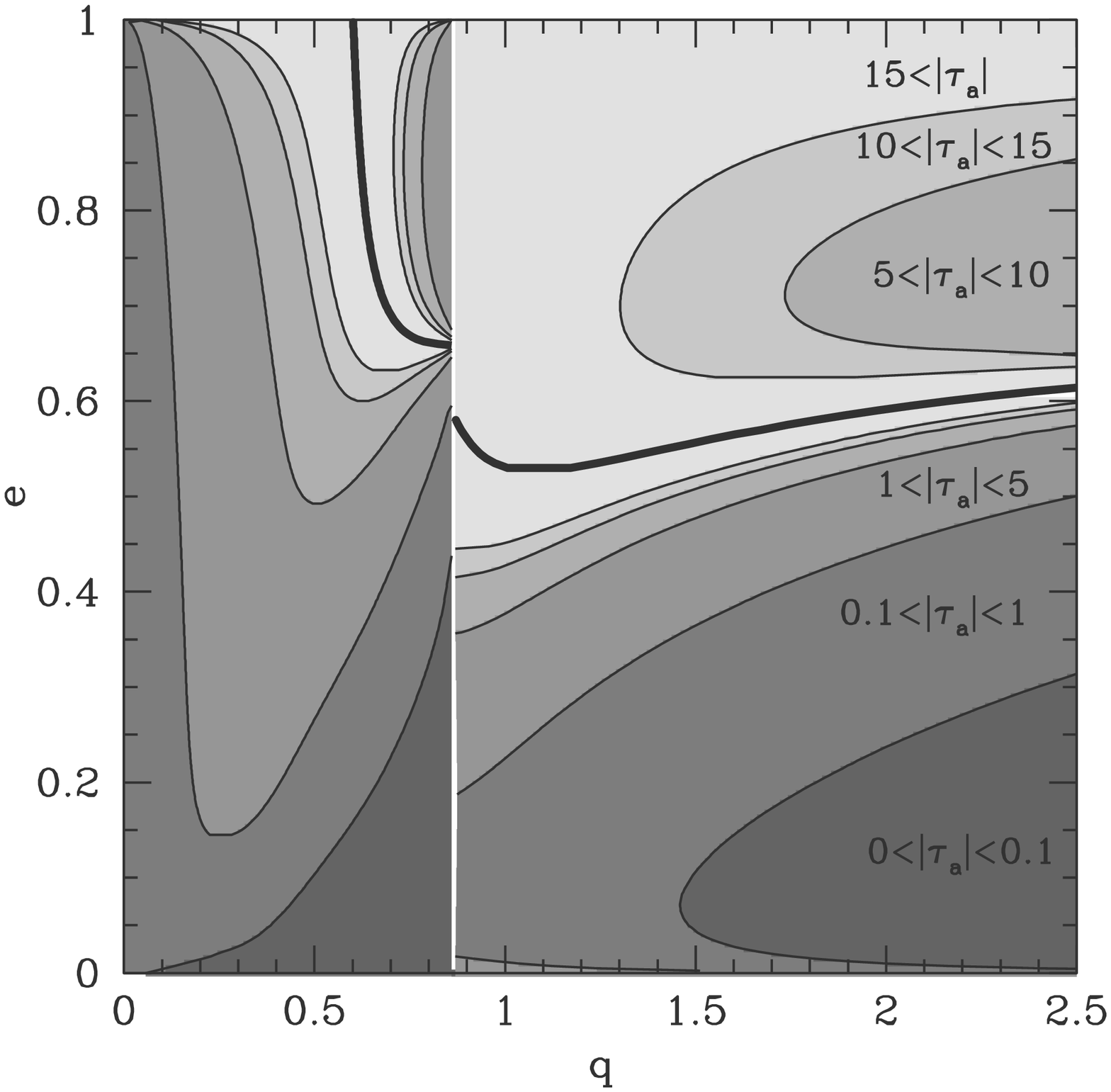}{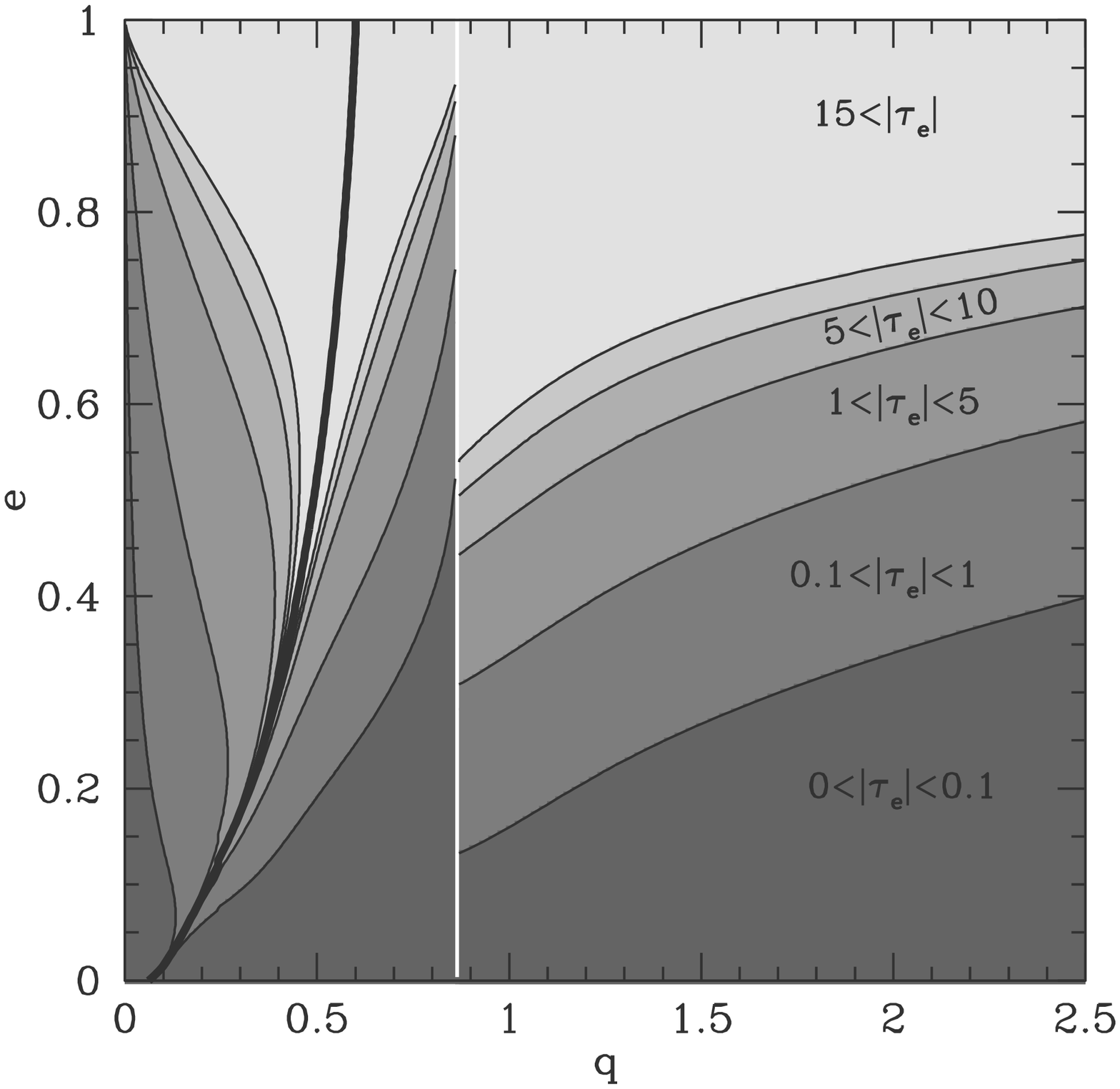}
\caption{Contour plots of the total orbital evolution timescales in the 
$(q,e$)-plane due to the combined effects of tidal dissipation and 
mass transfer for the same set of assumptions adopted in 
figure~~\ref{fig-tot_timescale}.  Timescales for the evolution of the 
semi-major axis are shown at left; timescales for the evolution of the 
orbital eccentricity are shown at right.  The different shades of gray 
designate regions of the ($q,e$) parameter space with timescales (in 
Gyr) in the ranges labeled in the plots.  The thick black line designates 
the transition point where the rate of change of the semi-major axis or 
orbital eccentricity changes from positive (to the left of the thick 
black line) to negative (to the right of the thick black line).  The 
vertical white line at $q\simeq0.87$ corresponds to a transition of the 
dominant tidal dissipation mechanism which is different for donor 
stars with convective envelopes ($M_1\lesssim1.25\,M_\sun$) than for 
donor stars with radiative envelopes ($M_1 \gtrsim1.25\,M_\sun$).}
\label{fig-tot_contour}
\end{figure*}

In figure~\ref{fig-tot_contour}, we show the total orbital evolution
timescale due to the sum of tidal and mass transfer effects as a contour
plot in the $(q,e)$-plane.  The thick black lines indicate the
transitions from positive (left of the thick black line) to negative
(right of the thick black line) rates of change of the semi-major axis
and eccentricity.  The white dividing line near $q\approx 0.87$
corresponds to the transition between tidal dissipation mechanisms in
stars with convective envelopes ($M_1\lesssim1.25\,{\rm M}_\sun$) and
stars with radiative envelopes ($M_1\gtrsim1.25\,{\rm M}_\sun$).  It
follows that there are large regions of parameter space where the
combined effects of mass transfer and tidal evolution do not rapidly
circularize the orbit.  In particular, for $q > 0.87$ and $e \ga 0.75$
orbital circularization always takes longer than 10\,Gyr, while for $q$
to the left of the thick black line the orbital eccentricity grows
rather than shrinks.  For a given $q$ left of the thick black line, the
timescales for eccentricity growth increase with increasing $e$ though,
so that there is no runaway eccentricity growth.  Hence, for small $q$,
mass transfer at the periastron of eccentric orbits may provide a means
for inducing non-negligible eccentricities in low-mass binary or
planetary systems.  The orbital semi-major axis, on the other hand,
always increases when $e \la 0.55$, but can increase as well as decrease
when $e \ga 0.55$, depending on the binary mass ratio $q$.  We recall
that both the tidal and mass transfer orbital evolution time scales
depend on the ratio of the donor's rotational angular velocity
$\Omega_1$ to the orbital angular velocity $\Omega_{{\rm orb},P}$ and that we have
set $\Omega_1/\Omega_{{\rm orb},P}=1$ in all figures shown.

\section{Concluding Remarks}

We developed a formalism to calculate the evolution of the
semi-major axis and orbital eccentricity due to mass transfer in
eccentric binaries, assuming conservation of total system mass and
orbital angular momentum. Adopting a delta-function mass-transfer
profile centered at the periastron of the binary orbit yields rates of
secular change of the orbital elements that are linearly proportional to
the magnitude $\dot{M}_0$ of the mass-transfer rate at the periastron.
For $\dot{M}_0 = 10^{-9}\,M_\sun\,{\rm yr}^{-1}$, this yields timescales
of orbital evolution ranging from a few Myr to a Hubble time or longer.
Depending on the initial binary mass ratio and orbital eccentricity, the
rates of secular change of the orbital semi-major axis and eccentricity
can be positive as well as negative, so these orbital elements can 
increase as well as decrease with time. 

Comparison of the timescales of orbital evolution due to mass transfer
with the timescales of orbital evolution due to tidal dissipation shows
that the effects can either be additive or competitive, depending on the
binary mass ratio, the orbital eccentricity, and the magnitude of the
mass-transfer rate at the periastron. Contrary to what is often assumed
in even the most state-of-the-art binary evolution and population
synthesis codes, tides do not always lead to rapid circularization
during the early stages of mass transfer. Thus, phases of episodic mass
transfer may occur at successive periastron passages and may persist for
long periods of time.  As a first approximation, the evolution of the
orbital semi-major axis and eccentricity due to mass transfer in
eccentric binaries can be incorporated into binary evolution and
population synthesis codes by means of equations~(\ref{eq-deltaa}) and
(\ref{eq-deltae}) in which the mass-transfer rate is approximated by a
delta-function of amplitude $\dot{M}_0$ centered at the periastron of
the binary orbit.

In future papers, we will relax the assumption of conservation of total
system mass and orbital angular momentum, and examine the effects of
non-conservative mass transfer on the orbital elements of eccentric
binaries. We also intend to study the onset of mass transfer in
eccentric binaries in more detail, adopting realistic mass-transfer
rates appropriate for atmospheric Roche-lobe overflow in interacting
binaries as discussed by \citet{R88}.  We will consider individual 
binary systems that are known to be eccentric and transferring mass 
during periastron passage, as well as populations of eccentric 
mass-transferring binaries and their descendants.


\appendix

\section{Equipotential Surfaces and the Inner Lagrangian point in
  Eccentric Binaries}
\label{sec-appa}
\label{appL1}

\begin{figure}
\plotone{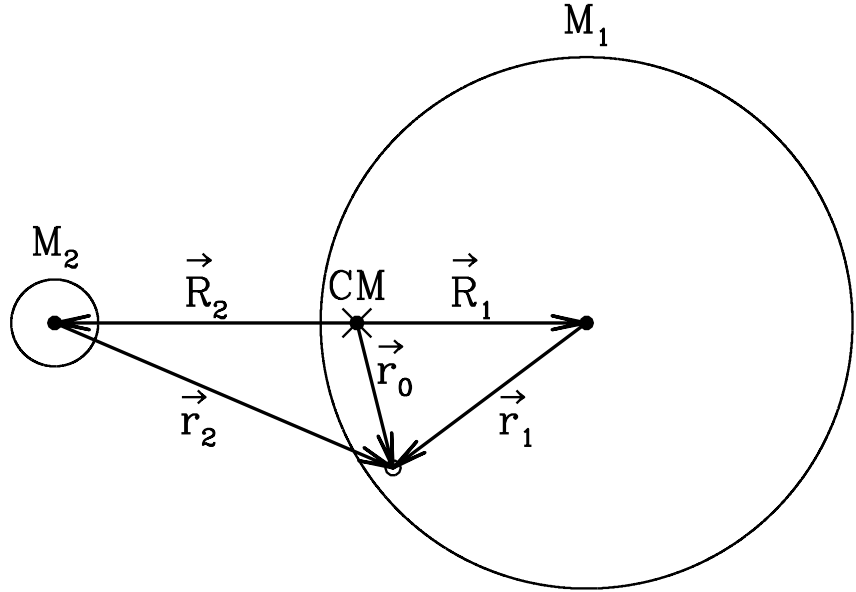}
\caption{Schematic diagram showing the vectors pertinent to the
  derivation of the equipotential surfaces for a non-synchronous,
  eccentric binary.  The center of mass of the system is shown
  as a cross near the center of the diagram and the center of mass of
  each star is shown as a filled circle at the star's center.  The 
  vectors $\vec{R}_1$ and $\vec{R_2}$ begin at the center of mass of 
  the system.}
\end{figure}

A crucial element in the description of mass transfer in any binary
system is the location of the inner Lagrangian point ($L_1$) through
which matter flows from the donor to the accretor.  While a solution
for the location of $L_1$ is not analytic, the case of circular orbits
with synchronized components can be approximated by the formula
\begin{equation}
X_{L_1} = 0.5 + 0.22 \log{q},  \label{eq-XL1cs}
\end{equation}
where $X_{L_1}$ is the distance of the $L_1$ point from the mass
center of the donor star in units of the distance between the stars,
and  $q=M_1/M_2$ is the mass ratio of the binary defined as the ratio
of the donor mass $M_1$ to the accretor mass $M_2$. For mass ratios in
the range $0.1 < q < 15$, this formula is valid to an accuracy of
better than 2\% \citep[e.g.][]{DR74}.  The above
formula has been generalized by \citet{PS76} to
include the effect of a non-synchronously rotating donor star: 
\begin{equation}
X_{L_1} = \left[ 0.53 - 0.03 \left( \frac{\Omega_1}{\Omega_{\rm orb}}
  \right)^2 \right] \left( 1 + \frac{4}{9}\log{q} \right). 
  \label{PS}
\end{equation}
Here, $\Omega_1$ is the rotational angular velocity of the donor, and
$\Omega_{\rm orb}$ the orbital angular velocity of the binary. For $0
< \Omega_1/\Omega_{\rm orb} < 1$ and $0.2 < q < 10$, the formula is
valid to an accuracy of better than 3\%. It is to be noted though that
\citet{PS76} derived the position of the $L_1$ point under the
assumption that the orbital and rotational periods of the binary and
its component stars are much longer than the dynamical timescale of
the donor. Despite the better than 3\% accuracy of the fit given by
equation~(\ref{PS}), the formula may therefore still break down because the
underlying formalism is no longer valid.

In this paper, it is necessary to generalize the formula for the
position of the $L_1$ point even further to account for a non-zero
eccentricity as well as non-synchronous rotation. For this purpose, we
determine the equipotential surfaces describing the shape of the
components of a non-synchronous, eccentric binary. Our procedure is a
generalization of the steps outlined by \citet{L63} for the
derivation of the equipotentials of a non-synchronous, circular binary.

Here, we consider an eccentric binary system where the stars are
considered to be centrally condensed and spherically symmetric, and
thus can be well described by Roche models of masses $M_1$ and $M_2$.
Their orbit is assumed to be Keplerian with semi-major axis $a$ and
eccentricity $e$. Star~1 is furthermore assumed to rotate uniformly
with a constant angular velocity $\vec{\Omega}_1$ parallel to the
orbital angular velocity $\vec{\Omega}_{\rm orb}$. We note that for an
unperturbed Keplerian orbit, the magnitude of $\vec{\Omega}_{\rm orb}$
is a function of time, while the direction and orientation remain
fixed in space.

As is customary, we determine the equipotential surfaces with respect
to a Cartesian coordinate frame $OXYZ$ with origin $O$ at the mass
center of star 1, and with the $Z$-axis pointing along
$\vec{\Omega}_1$. The $X$- and $Y$-axes are co-rotating with the star
at angular velocity $\Omega_1$. Since non-synchronously rotating
binary components are inevitably subjected to time-dependent tides
invoked by their companion, the mass elements will oscillate with
frequencies determined by the difference between the rotational and
orbital frequencies. Here, we neglect these tidally induced
oscillations as well as any other type of bulk motion of matter due
to, e.g., convection. This approximation is valid as long as the
characteristic timescale associated with the motion of the mass
elements is sufficiently long compared to the star's dynamical
timescale (more extended discussions on the validity of the
approximation can be found in \citet{L63}, \citet{S78}, and 
\citet{SWK07}).  Star 1 is then completely stationary with respect to 
the co-rotating frame of reference. 

With respect to the $OXYZ$ frame, the equation of
motion of a mass element at the surface of star 1 is given by
\begin{equation}
\ddot{\vec{r}}_1 = \ddot{\vec{r}}_0-\ddot{\vec{R}}_1 - \vec{\Omega}_1
\times \left( \vec{\Omega}_1 \times \vec{r}_1 \right) - 2\,
\vec{\Omega}_1 \times \dot{\vec{r}}_1,  
\label{eq-r1}
\end{equation}
where $\vec{r}_0$ and $\vec{r}_1$ are the position vectors of the mass
element with respect to the mass center of the binary and 
star 1, respectively, and $\vec{R}_1$ is the position vector of the center
of mass of star 1 with respect to the center of mass of the binary, and  
$-\vec{\Omega}_1 \times ( \vec{\Omega}_1 \times \vec{r}_1 )$ and $-
2\, \vec{\Omega}_1 \times \dot{\vec{r}}_1$ are the centrifugal and
Coriolis acceleration with respect to the rotating coordinate frame. 

Under the assumption that, in an inertial frame of reference, the only
forces acting on the considered mass element are those resulting from
the pressure gradients in star 1 and the gravitational attractions of
star 1 and its companion, the acceleration of the mass element with
respect to the binary's center of mass is given by
\begin{equation}
\ddot{\vec{r}}_0 = -\frac{1}{\rho}\, \vec{\nabla} P - \vec{\nabla} 
\left( -G\, \frac{M_1}{|\vec{r}_1|} - G\, \frac{M_2}{|\vec{r}_2|} 
\right),
\label{eq-r0}
\end{equation}
where $G$ is the Newtonian constant of gravitation, $\rho$ the mass
density, $P$ the pressure, and $\vec{r}_2$ the position vector of the
mass element with respect to the mass center of star 2.  The gradient in 
equation~(\ref{eq-r0}) and subsequent equations in this Appendix are 
taken with respect to $X$, $Y$, and $Z$.

The acceleration of the center of mass of star 1 with respect to the
binary mass center, on the other hand, is given by
\begin{equation}
\ddot{\vec{R}}_1 = -\frac{GM_2}{D^2} \frac{\vec{R}_1}{|\vec{R}_1|}, 
\label{eq-R}
\end{equation}
where $D(t)$ (simplified to $D$ in what follows) is the time-dependent
distance between star 1 and star 2.  

Substituting equations~(\ref{eq-r0}) and~(\ref{eq-R}) into
equation~(\ref{eq-r1}), we obtain 
\begin{equation}
\ddot{\vec{r}}_1 = -\frac{1}{\rho} \vec{\nabla} P - \vec{\nabla} \left( 
-G\,
\frac{M_1}{|\vec{r}_1|} - G\, \frac{M_2}{|\vec{r}_2|} \right) +
\frac{GM_2}{D^2} \frac{\vec{R}_1}{|\vec{R}_1|} - \vec{\Omega}_1
\times ( \vec{\Omega}_1 \times \vec{r}_1 ) - 2\, \vec{\Omega}_1 \times
\dot{\vec{r}}_1.
\label{eq-r1.2}
\end{equation}

When the $X$-axis coincides with the line connecting the mass centers
of the two stars, the third and fourth term in the right-hand member
of equation~(\ref{eq-r1.2}) can be written as the gradient of a potential
function as  
\begin{equation}
\frac{GM_2}{D^2} \frac{\vec{R}_1}{|\vec{R}_1|} = 
 - \vec{\nabla} \left( \frac{GM_2}{D^2} X \right),
\end{equation}
\begin{equation}
\vec{\Omega}_1\times ( \vec{\Omega}_1 \times \vec{r}_1 ) = -\vec{\nabla}
\left[ \frac{1}{2} |\vec{\Omega}_1|^2 \left( X^2 + Y^2 \right) \right],
\end{equation}
where $X$ and $Y$ are the Cartesian coordinates of the mass element
under consideration. With these transformations, equation~(\ref{eq-r1.2})
becomes 
\begin{equation}
\ddot{\vec{r}}_1 = -\frac{1}{\rho} \vec{\nabla} P - \vec{\nabla} V_1 - 
2\, \vec{\Omega}_1 \times \dot{\vec{r}}_1,  \label{a8}
\end{equation}
where
\begin{equation}
V_1 = -G \frac{M_1}{|\vec{r}_1|} - G \frac{M_2}{|\vec{r}_2|} -
\frac{1}{2} |\vec{\Omega}_1|^2 (X^2 + Y^2) + \frac{GM_2}{D^2} X.
\end{equation}

Since star 1 is assumed to be static in the rotating frame,
$\dot{\vec{r}}_1 = \ddot{\vec{r}}_1 = 0$, so that equation~(\ref{a8})
reduces to
\begin{equation}
\vec{\nabla} P = - \rho \vec{\nabla} V_1.
\end{equation}
This equation governs the instantaneous shape of the equipotential 
surfaces at the instant when the $X$-axis coincides with
the line connecting the mass centers of the two stars. However, since
this instant is entirely arbitrary, the equation is generally valid
and can be used to determine the shape of the star at any phase of the
orbit, with appropriate re-definition of the $X$-axis at every
instant. A similar equation was derived by \citet{A76} and
\citet{W79}.

Next, we look for the critical equipotential surface for
which matter starts to flow from star 1 to star 2 through the inner
Lagrangian point $L_1$. For circular binaries with synchronously
rotating component stars, the $L_1$ point is always located on the
line connecting the mass centers of the two stars. \citet{MW83} have
shown that non-synchronous rotation combined with a misalignment 
between the spin axis of star 1 and the orbital plane may cause the 
$L_1$ point to oscillate in the $Z$-direction.  As a first 
approximation, we here neglect
these oscillations and assume the $L_1$ point to be located on the
line connecting the mass centers of the two stars. The position of the
$L_1$ point is then obtained by setting $dV_1/dX=0$. The resulting
equation for the position of $L_1$ is
\begin{equation}
G \frac{M_1}{X^2} - G \frac{M_2}{(X-D)^2} -
  |\vec{\Omega}_1|^2 X + G \frac{M_2}{D^2} = 0,
\end{equation}
which can be rewritten in dimensionless form as
\begin{equation}
\frac{q}{X_{L_1}^2} - \frac{1}{(X_{L_1}-1)^2} 
  - f^2 X_{L_1} (1+q)(1+e) + 1 = 0. 
  \label{XL}
\end{equation}
Here, $X_{L_1}=X/D$ is the position of the $L_1$ point on the $X$-axis
in units of the distance between the two stars, $q=M_1/M_2$ is the
binary mass ratio, $f=\Omega_1/\Omega_{{\rm orb},P}$ is the ratio of the star 1's
rotational angular velocity to the orbital angular velocity at
periastron, and $\Omega_{{\rm orb},P}^2=G(M_1+M_2)(1+e)/D_p^3$. In the latter
expression, $D_P=a(1-e)$ is the distance between the stars at
periastron. Note that in the particular case of a circular orbit,
equation~(\ref{XL}) reduces to the equation for the position of the $L_1$
point derived by \citet{K63} and \citet{PS76}.

\begin{figure}
\plotone{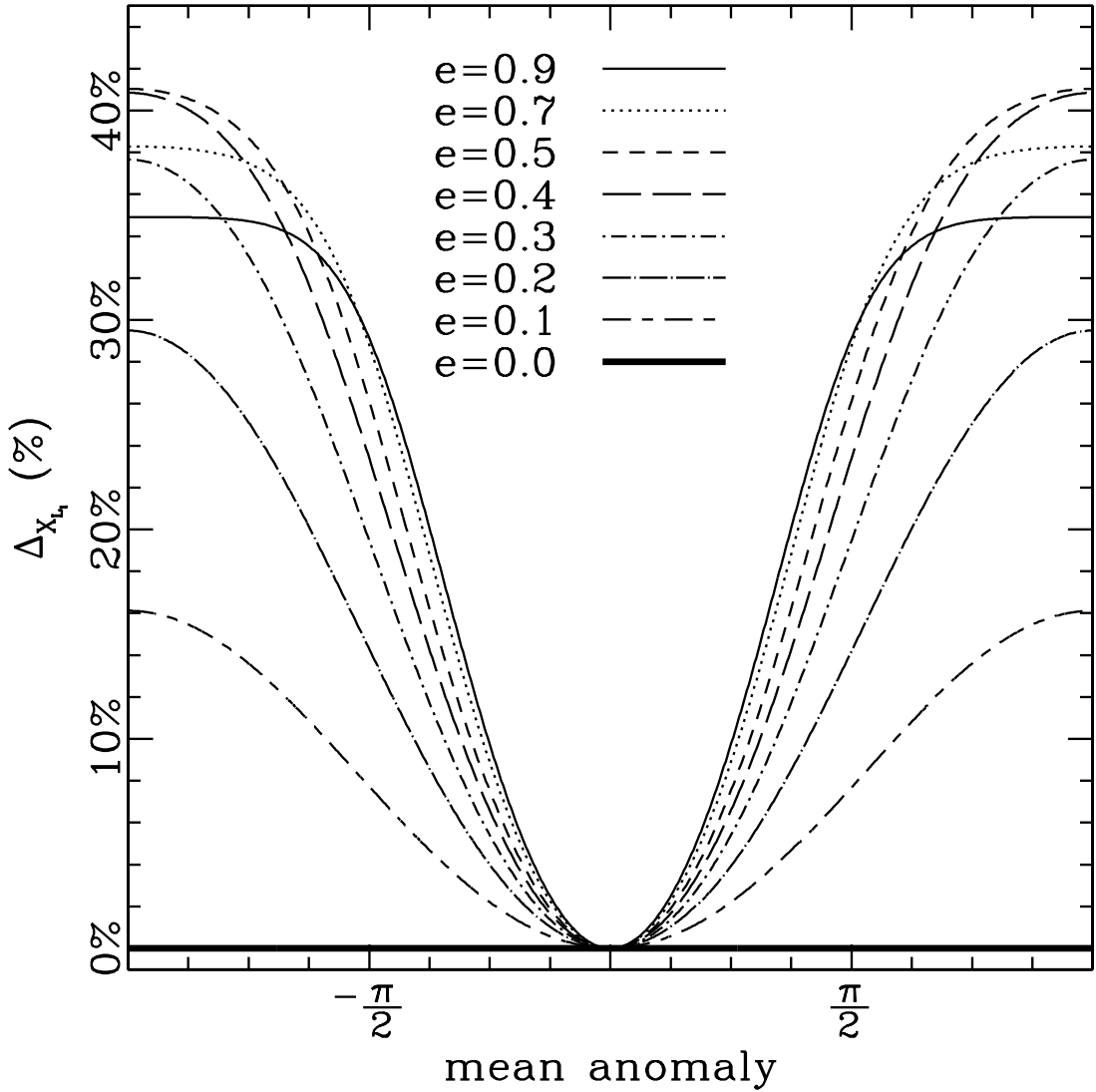}
\caption{The relative change in the position of the $L_1$ point as a
  function of true anomaly $\nu$, for $q=1$, $f=1$, 
and a range of
  eccentricities.}
\label{fig-l1}
\end{figure}

Equation~(\ref{XL}) must be solved numerically for the location of
$L_1$.  Since we have made no explicit assumptions about the relative
location of the two star in the orbit, and we have let the distance
between the stars $D=a(1-e^2)/(1+e\cos{\nu})$, with $\nu$ the true
anomaly, be an explicit function of time, this equation characterizes
the position of the $L_1$ point at any position in the orbit.
Figure~\ref{fig-l1} shows the relative change
\begin{equation}
\Delta_{X_{L_1}}= \frac{X_{L_1}(\nu)-X_{L_1}(\nu=0)}{X_{L_1}(\nu=0)}
\end{equation}
in the position of the
$L_1$ point as a function of the true anomaly for $q=1$, $f=1$,
and a variety of eccentricities.  The position of the $L_1$ point over
the course of a single orbit varies by less than 15\% when $e \la
0.1$, and by 30--40\% when $e \ga 0.2$. At apastron, $\Delta_{X_{L1}}$
increases with increasing eccentricity when $e \la 0.5$, and decreases
with increasing eccentricity when $e \ga 0.5$. This turn-around at $e
\simeq 0.5$ is caused by the increasing deviations from synchronous
rotation at apastron with increasing orbital eccentricity.

We have also derived a fitting formula for the position of the $L_1$
point at the periastron of the binary orbit as a function of $q$, $e$,
and $f$:
\begin{equation}
X_{L_1} = 0.529+0.231\log{q} - f^2(0.031 + 0.025 e)(1+0.4\log{q}).
\label{XL1fit}
\end{equation}
This fitting formula is accurate to better than $4\,\%$ for mass
ratios $0.1 \le q \le 10$, eccentricities $0 \le e \le 0.9$, and
ratios of stellar rotational angular velocities to orbital angular
velocities at periastron $0 \le f \le 1.5$. In the limiting case of a
circular, synchronized binary, equation~(\ref{XL1fit}) reduces to
equation~(\ref{eq-XL1cs}), while for a circular, non-synchronized binary,
equation~(\ref{XL1fit}) reduces to equation~(\ref{PS}).  A detailed discussion of 
the properties of all five Lagrangian points and the volume of the 
critical Roche Lobe in eccentric binaries is presented in \citet{SWK07}


\section{Orbital Evolution Due to Instantaneous Mass Transfer Between 
Two Point Masses}
\setcounter{figure}{0}
\label{sec-appc}

When the components of a mass-transferring binary are treated as point
masses and mass transfer is assumed to be instantaneous, an alternative
derivation of the equations governing the secular change of the orbital
semi-major axis and eccentricity may be obtained from the equations for
the specific orbital energy and angular momentum.  This method is
commonly used to relate the post- to pre-supernova orbital parameters of
binaries in which one of the components undergoes an instantaneous
supernova explosion \citep[e.g.][]{H83, BP95, K96}. As in
Section~\ref{orbevtim}, we assume conservation of total system mass and
describe the mass transfer by a delta function mass-transfer rate
centered at the periastron of the binary orbit [see equation~(\ref{eq-del})].

The relative orbital velocity, $V_{\rm rel}$, of two stars with masses
$M_1$ (the donor star) and $M_2$ (the accreting star) in a binary with
orbital semi-major axis $a$ and eccentricity $e$ is given by
\begin{equation}
V_{\rm rel}^2 = G(M_1+M_2) \left( \frac{2}{|\vec{r}|} - \frac{1}{a} 
\right),
\label{eq-C-V}
\end{equation}
where $G$ is the Newtonian constant of gravitation, and $\vec{r}$ is the
relative position vector of the accretor with respect to the donor. 
Under the assumption that mass elements are transferred instantaneously
between the two stars, the distance $|\vec{r}|$ is the same right before
and right after mass transfer, so that the rate of change of the orbital
semi-major axis due to instantaneous mass transfer is given by
\begin{equation}
\frac{da}{dt} = \frac{2a^2V_{\rm rel}}{G(M_1+M_2)}\frac{dV_{\rm rel}}{dt}.
\label{eq-C-dadt}
\end{equation}

Next, the specific orbital angular momentum of the binary written in 
terms of the binary component masses and orbital elements is given by
\begin{equation}
|\vec{r}\times\vec{V}_{\rm rel}|^2 = G(M_1+M_2)a(1-e^2).
\label{eq-C-J}
\end{equation}
Noting that $|\vec{r}\times\vec{V}_{\rm rel}|=|\vec{r}||\vec{V}_{\rm
rel}|$ at periastron, and proceeding in a similar way as for the
derivation of equation~(\ref{eq-C-dadt}), we derive the rate of change
of the orbital eccentricity due to instantaneous mass transfer to be
\begin{equation}
\frac{de}{dt} = \frac{2a(1-e)|\vec{V}_{\rm 
rel}|}{G(M_1+M_2)}\frac{d|\vec{V}_{\rm rel}|}{dt}. 
\label{eq-C-dedt}
\end{equation}

Finally, we eliminate $V_{\rm rel}$ from equations~(\ref{eq-C-dadt})
and~(\ref{eq-C-dedt}) to obtain equations for the rates of changes of
the orbital semi-major axis and eccentricity in terms of the binary
orbital elements and component masses. For this purpose, we use the
conservation of linear momentum to write the rate of change of the
relative orbital velocity due to mass-transfer as
\begin{equation}
\frac{d\vec{V}_{\rm rel}}{dt} = \frac{\dot{M}_2}{M_2}\, \vec{v}_{\delta 
M_2} - \frac{\dot{M}_1}{M_1}\, \vec{v}_{\delta M_1},
\label{eq-C-vrelvec}
\end{equation}
where $\vec{v}_{\delta M_1}$ and $\vec{v}_{\delta M_2}$ are the relative
velocities of the transferred mass elements with respect to the mass
center of the donor and accretor, respectively.  To obtain an equation 
for the change in the magnitude of $\vec{V}_{\rm rel}$, we write the 
vector components of $\vec{V}_{\rm rel}$, $\vec{v}_{\delta M_1}$, and 
$\vec{v}_{\delta M_2}$ with respect to the $\hat{x}$, $\hat{y}$, and 
$\hat{z}$ unit vectors introduced in Section~\ref{relmot}, and take the 
dot product of equation~(\ref{eq-C-vrelvec}) with $\vec{V}_{\rm rel}$. 
Since, $\vec{V}_{\rm rel} = V_{\rm rel}\, \hat{y}$ at the periastron of 
the binary orbit, it follows that
\begin{equation}
\frac{dV_{\rm rel}}{dt} = -\frac{\dot{M}_1}{M_1} \left( qv_{\delta 
M_{2,y}} + v_{\delta M_{1,y}} \right).
\label{eq-C-vrel}
\end{equation}
Assuming conservation of orbital angular momentum it then follows from
equations~(\ref{eq-Jdot}) and~(\ref{eq-C-dadt})--(\ref{eq-C-vrel}) that
\begin{equation}
qv_{\delta M_2,y} + v_{\delta M_1,y}= na(1-q)\left(\frac{1+e}{1-e} 
\right)^{1/2},
\label{eq-C-vdm2}
\end{equation}
where we have used Kepler's 3rd law and that $|\vec{V}_{\rm rel}| = 
\left(G(M_1+M_2)/a\right)^{1/2}\left((1+e)/(1-e)\right)^{1/2}$ at 
the periastron of the binary orbit.  This equation is identical to 
equation~(\ref{eq-vr}) in the limiting
case where $|\vec{r}_{A_1}| \rightarrow 0$ and $|\vec{r}_{A_2}| 
\rightarrow 0$.  Substituting equations~(eq-C-vdm2) and (eq-C-vrel) into 
equations~(\ref{eq-C-dadt}) and
(\ref{eq-C-dedt}) and averaging over the orbital period, we find
\begin{equation}
\left< \frac{da}{dt} \right>_{\rm sec} = 
\frac{a}{\pi}\frac{\dot{M}_0}{M_1}\left( 1-e^2 
\right)^{1/2}(q-1)
\label{eq-C-dadtsec}
\end{equation}
\begin{equation}
\left< \frac{de}{dt} \right>_{\rm sec}= 
\frac{1}{\pi}\frac{\dot{M}_0}{M_1}(1-e^2)^{1/2}(1-e)(q-1).
\label{eq-C-dedtsec}
\end{equation}
These equations are identical to equations~(\ref{eq-dadtsec})
and(\ref{eq-dedtsec}) in the limiting case where
$|\vec{r}_{A_1}|\rightarrow0$ and $|\vec{r}_{A_2}|\rightarrow0$.

\acknowledgements

We are grateful to an anonymous referee for insightful comments which
helped to improve the paper, as well as to Ronald Taam for useful
discussions, and to Richard O'Shaughnessy for discussions regarding
methods used to fit the position of the $L_1$ point in eccentric,
non-synchronous binaries. This work is partially supported by a NASA
Graduate Fellowship (NNG04GP04H/S1) to J.S., and a NSF CAREER Award
(AST-0449558), a Packard Fellowship in Science and Engineering, and a
NASA ATP Award (NAG5-13236) to V.K.


\end{document}